\documentclass[aps,pra,showpacs,twocolumn,superscriptaddress]{revtex4-1}
\usepackage{amsfonts,amsmath,amssymb,graphicx}
\usepackage[usenames,dvipsnames]{xcolor}
\usepackage{pdfpages}
\usepackage{footmisc}
\usepackage[normalem]{ulem}
\usepackage{color}

\begin{document}

\title{When polarons meet polaritons: Exciton-vibration interactions in organic molecules strongly coupled to confined light fields}

\author{Ning Wu}
\email{wun1985@gmail.com; \emph{Present Address}: Center for Quantum Technology Research, School of Physics, Beijing Institute of Technology, Bejing 100081, China}
\affiliation{Departamento de F\'isica Te\'orica de la Materia Condensada and Condensed Matter Physics Center (IFIMAC), Universidad Aut\'onoma de Madrid, E-28049 Madrid, Spain}
\author{Johannes Feist}
\email{johannes.feist@uam.es}
\affiliation{Departamento de F\'isica Te\'orica de la Materia Condensada and Condensed Matter Physics Center (IFIMAC), Universidad Aut\'onoma de Madrid, E-28049 Madrid, Spain}
\author{Francisco J. Garcia-Vidal}
\email{fj.garcia@uam.es}
\affiliation{Departamento de F\'isica Te\'orica de la Materia Condensada and Condensed Matter Physics Center (IFIMAC), Universidad Aut\'onoma de Madrid, E-28049 Madrid, Spain}
\affiliation{Donostia International Physics Center (DIPC), E-20018 Donostia/San Sebasti\'an, Spain}

\begin{abstract}
We present a microscopic semi-analytical theory for the description of organic molecules interacting strongly with a cavity mode. Exciton-vibration coupling within the molecule and exciton-cavity interaction are treated on an equal footing by employing a temperature-dependent variational approach. The interplay between strong exciton-vibration coupling and strong exciton-cavity coupling gives rise to a hybrid ground state, which we refer to as the lower polaron polariton. Explicit expressions for the ground-state wave function, the zero-temperature quasiparticle weight of the lower polaron polariton, the photoluminescence line strength, and the mean number of vibrational quanta are obtained in terms of the optimal variational parameters. The dependence of these quantities upon the exciton-cavity coupling strength reveals that strong cavity coupling leads to an enhanced vibrational dressing of the cavity mode, and at the same time a vibrational decoupling of the dark excitons, which in turn results in a lower polaron polariton resembling a single-mode dressed bare lower polariton in the strong-coupling regime. Thermal effects on several observables are briefly discussed.
\end{abstract}


\maketitle

\section{Introduction}\label{I}
\par Over the last couple of decades, there has been a renewed interest in organic molecular materials because of their high relevance to organic light-emitting diodes~\cite{LED1,LED2}, organic lasers~\cite{OL}, organic solar cells~\cite{OSC}, organic field-effect transistors~\cite{OFET}, and natural/artificial light-harvesting systems~\cite{LH1,LH2}. Organic materials are also ideal systems to achieve strong coupling with confined light fields due to their large dipole moments and possible high molecular densities. The strong-coupling regime is entered when the coherent energy exchange between emitters and light modes becomes faster than decay and decoherence processes in either constituent. This leads to the formation of two polariton modes, i.e., hybrid eigenstates that have mixed light-matter character, separated by the Rabi splitting. Strong coupling of organic molecules has been studied in a wide variety of photonic systems, among them dielectric microcavities~\cite{Agvich1997,Lidzey1998,Lidzey1999,Holmes2004,Kena-Cohen2008,2015review}, metallic microcavities~\cite{Schwartz2011,Hutchison2012}, plasmonic modes on flat~\cite{Bellessa2004,Hakala2009,Berrier2011} and holey surfaces~\cite{Dintinger2005,Vasa2008}, and nanoparticle arrays supporting surface lattice resonances~\cite{Rodriguez2013,Vakevainen2014}. Additionally, the strong field confinement in plasmonic systems also allows strong coupling with localized surface plasmon resonances~\cite{Wurtz2007,Zengin2015,Eizner2015}, even down to the single-molecule level~\cite{2016Nature}. The Rabi splitting in these vastly different systems is all quite similar, with typical values of hundreds of meV and reaching up to more than $1$~eV~\cite{Schwartz2011,Kena-Cohen2013}.

\par Despite the fact that organic molecules offer an excellent platform to enter the strong light-matter interaction regime, in most theoretical descriptions they are often modeled as simple two-level systems whose coupling to the cavity field forms the usual hybrid light-matter excitations called polaritons. Very recently, there have appeared a few theoretical works explicitly including intramolecular vibrations (or optical phonons)~\cite{Keeling2014,Spano2015,Spano2016,PRX,Keeling2016,KeelingNew}, which were first suggested by Holstein~\cite{Holstein} to play an essential role in the understanding of charge-carrier transport mechanisms in organic molecular crystals. Along this line, very recently, Spano~\cite{Spano2015} studied the effects of exciton-cavity coupling on the static zero-temperature properties of J-aggregates using numerical diagonalization of the Frenkel-Holstein model in a truncated subspace with a finite number of vibrational quanta.
\par In this work, by using a full quantum model built upon the Holstein Hamiltonian, we explicitly treat exciton-vibration coupling and exciton-cavity coupling on an equal footing. Such a model is suitable for describing a variety of low-dimensional organic materials interacting with cavity fields, including J-aggregates~\cite{Jagg} and light-harvesting complexes~\cite{LHII0,LHII}, among others. The static properties of the system are then studied by a generalized temperature-dependent variational Merrifield transformation that includes the vibrational dressing of both the exciton and the cavity mode, even though the latter does not interact with the vibrations directly. Originally proposed by Merrifield~\cite{Merrifield} and later developed by Silbey and co-workers~\cite{Silbey1977,Cheng2008}, the variational polaron transformation approach provides a convenient and accurate description of both static (e.g., ground-state) and dynamical (e.g., finite-temperature charge-carrier mobility) properties of organic molecular systems, even in the intermediate exciton-vibration coupling regime. Recently, polaronlike transformations have also been used in the study of electron-phonon interaction effects in quantum-dot-cavity systems~\cite{Nazir1,Nazir2,Roy2011,Mork2012}.

\par By taking the additional vibrational dressing of the cavity mode into account, the temperature-dependent variational canonical transformation approach employed in the present work provides an intuitive way to capture the main physics of the system, and the analytical results enable transparent physical interpretations of the observed phenomena. It also gives a natural way to study static and dynamical properties of organic microcavities at finite temperatures for a wide range of parameters. As we will show, the transformation not only yields renormalization on the exciton hopping integral, the exciton-cavity coupling strength, and the cavity frequency, but also induces an effective exciton-vibration coupling in the transformed frame. At zero temperature, we benchmark our method by a generalized Toyozawa ansatz and show that both approaches give accurate results for the ground state for a wide range of parameters. As a semi-analytical method, we are allowed to derive explicit forms of the ground-state energy and ground-state wave function. It turns out that the ground state is a highly entangled state containing both polaronlike and polaritonlike structures. We thus call the corresponding quasiparticle a lower polaron polariton (LPP) to distinguish from the usual lower exciton polariton (LP)~\cite{LP1992}. For fixed material parameters, we calculate the quasiparticle weight, the photoluminescence line strength, and the mean number of vibrational quanta as functions of the exciton-cavity coupling strength. The variation of these quantities with increasing exciton-cavity coupling indicates that strong exciton-cavity coupling induces a reduction of vibrational dressing of the excitons, but an enhancement of vibrational dressing of the cavity state. We finally study the thermal effects on the above observables by using the zero-order density matrix of the system. In the strong exciton-cavity coupling regime, the system develops a large energy gap between the lowest dark exciton and the LPP state, yielding an almost temperature-independent behavior below a crossover temperature.

\par The rest of the paper is structured as follows. In Sec.~\ref{II}, we introduce the model and describe our generalized Merrifield method in detail. In Sec.~\ref{III}, we present the results for the ground-state properties. Expressions for the ground-state wave function and the quasiparticle weight are given. Section~\ref{IV} is devoted to the calculation of several observables including the photoluminescence line strength and the mean number of vibrational quanta at both zero-temperature and finite temperatures. Conclusions are drawn in Sec.~\ref{V}.
\section{Model and methodology}\label{II}
\subsection{Hamiltonian}
\par A typical organic microcavity setup consists of layer-structured organic materials sandwiched between two dielectric mirrors that form the microcavity~\cite{AFM}. Most recently, strongly coupled organic microcavities with single/few molecules have been realized experimentally, where the volume of the microcavity can be scaled to less than $40$ cubic nanometers by employing a nanoparticle-on-mirror geometry~\cite{2016Nature}. For simplicity, we consider an organic microcavity composed of a single one-dimensional organic molecule located in a single-mode cavity.  The single organic molecule is assumed to consist of $N$ chromophores. Such a system is described by the Hamiltonian
\begin{eqnarray}\label{H}
H&=&H_{\rm{mat}}+H_{\rm{c}}+H_{\rm{e-c}},\nonumber\\
H_{\rm{mat}}&=&H_{\rm{e}}+H_{\rm{v}}+H_{\rm{e-v}},\nonumber\\
H_{\rm{e}}&=&\sum_n \varepsilon_na^\dag_na_n+\sum_{n\neq m}J_{nm}a^\dag_na_m,~J_{nm}=J_{mn},\nonumber\\
H_{\rm{v}}&=&\sum_n\omega_n b^\dag_nb_n,~H_{\rm{c}}=\omega_cc^\dag c,\nonumber\\
H_{\rm{e-v}}&=&\sum_n \lambda_n \omega_n a^\dag_na_n(b_n+b^\dag_n),\nonumber\\
H_{\rm{e-c}}&=&g\sum_n(a^\dag_n c+c^\dag a_n).
\end{eqnarray}

\par The material part $H_{\rm{mat}}$ of $H$ is the Holstein Hamiltonian that describes the organic molecule with intramolecular vibrations. In principle, the molecule also interacts with the continuous phonon modes from its surrounding environment. Usually, such molecule-phonon coupling is weak, and we henceforth neglect the continuous phonon bath for the sake of simplicity. Here, $a^\dag_n$ creates an exciton state $|n\rangle$ on site $n$ with on-site energy $\varepsilon_n$, and $J_{mn}$ is the hopping matrix element between two distinct sites $m$ and $n$. The intra-molecular vibrational mode on site $n$ with frequency $\omega_n$ is created by the boson creation operator $b^\dag_n$. $H_{\rm{e-v}}$ is the linear exciton-vibration coupling with strength measured by the Huang-Rhys factor $\lambda^2_n$. The radiation part is described by $H_{\rm{c}}$ with photon creation operator $c^\dag$ and cavity frequency $\omega_{\rm{c}}$. The last term in \ref{H} represents the uniform exciton-cavity interaction with interaction strength $g$. Here, we have employed the rotating wave approximation (RWA) such that no counter-rotating terms are present and $H_{\rm e-c}$ conserves the total number of excitations. This approximation is valid provided the ultrastrong-coupling regime is not reached, i.e., as long as the Rabi splitting is significantly smaller than the excitation energies $\varepsilon_n$, $\omega_c$ (see Ref.~\cite{Keeling2016} for a discussion of possible effects caused by the breakdown of the RWA).

\par In this work, we will consider one-dimensional molecules with uniform on-site energies and nearest-neighbor electronic couplings, i.e., we set $\varepsilon_n=\varepsilon_0$ and $J_{mn}=J\delta_{m,n\pm1}$. Important examples include linear J-aggregates~\cite{Jagg} and the light-harvesting complex II with a ring-like structure~\cite{LHII0,LHII}. We assume periodic boundary conditions in the former case. We have checked that typical amounts of static disorder and inhomogeneous broadening do not significantly affect the results presented here. For simplicity, the vibrational modes are modeled by Einstein oscillators with a single frequency $\omega_n=\omega_0$ and uniform exciton-vibration coupling $\lambda_n=\lambda$. In the following, we will restrict ourselves to the single-excitation subspace with $\sum_n a^\dag_n a_n + c^\dag c=1$, such that (within the RWA) we can truncate the number of cavity photons to be, at most, one. In turn, we can write $a^\dag_n=|n\rangle\langle \rm{vac}|$ and $c^\dag=|c\rangle\langle \rm{vac}|$, where $|\rm{vac}\rangle$ is the common vacuum of all the annihilation operators appearing in \ref{H}, hence an eigenstate of $H$ with vanishing energy. Note that in the absence of the vibrations and phonons, the excitonic and cavity part of $H$ resembles an interacting central spin model with spins $1/2$~\cite{PRA2014}.

\par The system is translationally invariant in its material part due to the periodic boundary conditions imposed, which allows us to work in the momentum space of the molecule through the Fourier transforms
 \begin{eqnarray}\label{FT}
a_n&=&\frac{1}{\sqrt{N}}\sum_k e^{ikn}a_k,~b_n=\frac{1}{\sqrt{N}}\sum_q e^{iqn}b_q.
\end{eqnarray}
We see that only the exciton state with zero momentum, $|k=0\rangle=a^\dag_{0}|\rm{vac}\rangle$, couples to the cavity field, so that the total crystal momentum
 \begin{eqnarray}\label{P-crys}
P_{\rm{tot}}=\sum_k ka^\dag_k a_k+\sum_q qb^\dag_q b_q
\end{eqnarray}
is still a good quantum number.
\subsection{The generalized Merrifield transformation}
\par In this work, in order to treat the exciton-vibration coupling and exciton-cavity coupling at finite temperatures on an equal footing, we employ an extended variational Merrifield transformation~\cite{Merrifield} determined by minimizing the Bogoliubov upper bound for the free energy. As demonstrated for the Holstein model in Ref.~\cite{Cheng2008}, and more recently in Ref.~\cite{CP2016}, these kinds of variational canonical transformation methods could offer an accurate description of both static properties (e.g., the ground state, the optical spectra, etc.) and dynamical properties (e.g., the exciton transport mechanisms) from intermediate to strong exciton-vibration coupling regimes.

\par When the cavity field is introduced, it couples only to the bright exciton, and there is no direct interaction between the cavity mode and the vibrations (though explicit cavity-vibration coupling has been considered in Refs.~\cite{CV0,CV1,CV2,CV3}). However, as we will show below, in the framework of the canonical transformation, the interplay of the light-matter interaction with the exciton-vibration coupling will induce an effective cavity-vibration coupling in the residue interaction in the Merrifield frame. It is straightforward to extend the present method to nonuniform or disordered systems.

\par To obtain an optimal zero-order representation of the Hamiltonian (\ref{H}) for a wide range of parameters, we propose the following generalized Merrifield transformation
\begin{eqnarray}\label{MT}
\tilde{H}&=&e^{\mathcal{S}}He^{-\mathcal{S}},\nonumber\\
\mathcal{S}&=&-\sum_n a^\dag_n a_nB_n-c^\dag c B_{\rm{c}},
\end{eqnarray}
with vibrational operators
\begin{eqnarray}\label{Bn}
B_n&=&\sum_l f_l (b_{n+l}-b^\dag_{n+l}),~B_{\rm{c}}=h\sum_l (b_l-b^\dag_l).
\end{eqnarray}

\par The variational parameters $\{f_l\}$ and $h$ are chosen to be real and are determined self-consistently by minimizing the free energy of the transformed system using Bogoliubov's inequality~\cite{feynman},
\begin{eqnarray}
F\leq F_0+\langle \mathcal{H}_1\rangle_{\mathcal{H}_0}
\end{eqnarray}
for a generic Hamiltonian $\mathcal{H}=\mathcal{H}_0+\mathcal{H}_1$, where $F$ and $F_0$ are the free energies of $\mathcal{H}$ and $\mathcal{H}_0$, respectively, and $\langle...\rangle_{\mathcal{H}_0}$ represents the thermal average over the canonical ensemble defined by $\mathcal{H}_0$.
\par Physically, the coefficient $f_l$ quantifies the degree of dressing of an exciton at site $n$ by the vibrational mode at site $n+l$, while $h$ measures the degree of dressing of the cavity photon by the vibrational mode on each excitonic site, though the cavity is not directly coupled to the vibrations. The usual small polaron transformation for the Holstein model corresponds to the case of $f_l=\delta_{l0}\lambda$ and $h=0$. By introducing the Fourier transforms of $\{f_{l}\}$,
\begin{eqnarray}
\tilde{f}_{q}=\sum_n e^{iqn}f_n,
\end{eqnarray}
the extended Merrifield generator can be written in the momentum space as
\begin{eqnarray}\label{S_mom}
\mathcal{S}&=&- \frac{1}{\sqrt{N}}\sum_{ kq}a^\dag_{k+q} a_{k}  \tilde{f}_q(b_{q}-b^\dag_{-q})-c^\dag c \sqrt{N}h(b_0-b^\dag_0),\nonumber\\
\end{eqnarray}
which clearly converses the total crystal momentum $P_{\rm{tot}}$ of the transformed states.

\par Besides the circular symmetry, the exciton-vibration system also holds inversion symmetry, which reduces the number of independent variational parameters from $N$ to $\frac{N}{2}+1$ ($\frac{N+1}{2}$), i.e., $\{f_0,f_1=f_{N-1},...,f_{\frac{N}{2}-1}=f_{\frac{N}{2}+1},f_{\frac{N}{2}}\}$ ($\{f_0,f_1=f_{N-1},...,f_{\frac{N-1}{2}}=f_{\frac{N+1}{2}}\}$) for even (odd) $N$~\cite{Cheng2008}. Using the tilde to indicate the Merrifield frame, the transformed Hamiltonian can be separated in a conventional way as
\begin{eqnarray}\label{Htilde}
\tilde{H}&=&\tilde{H}_{\rm{S}}+\tilde{V}+H_{\rm{v}},
\end{eqnarray}
where the system part reads
\begin{eqnarray}\label{H_Stilde}
\tilde{H}_{\rm{S}}&=&\sum_{k\neq0}E_k a^\dag_ka_k+[\varepsilon_0 a^\dag_0a_0+\tilde{g}\sqrt{N } (  a^\dag_0 c+ c^\dag a_0 )+\tilde{\omega}_{\rm{c}}c^\dag c].\nonumber\\
\end{eqnarray}
Here,
\begin{eqnarray}\label{bareband}
E_k=\varepsilon_0+\omega_0(\sum_m f^2_m-2\lambda f_0)+2\tilde{J}\cos k
\end{eqnarray}
is the vibrationally renormalized exciton dispersion and we have introduced the renormalized parameters
\begin{eqnarray}\label{tilde_para}
\tilde{J}=J\Theta_1,~\tilde{g}=g\Theta,~\tilde{\omega}_{\rm{c}}=\omega_{\rm{c}}+Nh^2\omega_0,
\end{eqnarray}
with
\begin{eqnarray}\label{TT1}
\Theta&=&\langle e^{B_{\rm{c}}-B_n}\rangle_{\rm{v}}=e^{-\frac{1}{2}\coth\frac{\beta\omega_0}{2}\sum_l(f_l-h)^2},\nonumber\\
\Theta_{|n-n'|}&=&\langle e^{B_n-B_{n'}}\rangle_{\rm{v}}=e^{-\frac{1}{2}\coth\frac{\beta\omega_0}{2}\sum_l(f_{l-n}-f_{l-n'})^2}.
\end{eqnarray}
where $\langle ...\rangle_{\rm{v}}=\frac{1}{Z_{\rm{v}}}\rm{Tr}_{\rm{v}}\{e^{-\beta H_{\rm{v}}}...\}$ is the thermal average with respect to the vibrational modes, with $\beta=1/k_BT$ the inverse temperature and $Z_{\rm{v}}=\rm{Tr}_{\rm{v}}e^{-\beta H_{\rm{v}}}$ the vibrational partition function. It is clear from Eqs.~(\ref{tilde_para}) and (\ref{TT1}) that the interaction with vibrations will \emph{decrease} both the effective hopping integral $J$ and the cavity coupling $g$, but will \emph{increase} the effective cavity frequency $\omega_{\rm{c}}$.
\par The residue interaction part is of the form
\begin{eqnarray}
\tilde{V} &=&\sum_{k_1,k_2}(P_{k_1k_2}+T_{k_1-k_2})a^\dag_{k_1}a_{k_2}+\sum_k(a^\dag_k cR_k+c^\dag a_k R^\dag_k)\nonumber\\
&&+\omega_0\sqrt{N}h c^\dag c(b_0+b^\dag_0),
\end{eqnarray}
where
\begin{eqnarray}\label{PQT}
P_{k_1k_2}&=& \frac{J}{N}\sum_{nm}\delta_{|m-n|,1}(e^{B_m-B_n}-\Theta_1) e^{-ik_1n+ik_2m},\nonumber\\
R_k&\equiv&\frac{g}{\sqrt{N}} \sum_n e^{-ikn} (e^{B_{\rm{c}}-B_n}-\Theta),\nonumber\\
T_{q}&\equiv& \frac{\omega_0}{\sqrt{N}}(\lambda-\tilde{f}_q)(b^\dag_{-q}+b_q),
\end{eqnarray}
are operators of vibrational degrees of freedom and satisfy $P_{k_1k_2}=P^\dag_{k_2k_1}$ and $T_q=T^\dag_{-q}$. We see that the extended Merrifield transformation leads to an \emph{effective cavity-vibration interaction} $\omega_0\sqrt{N}h c^\dag c(b_0+b^\dag_0)$ with the zero-momentum vibrational mode.

\par By examining the zero-order system Hamiltonian $\tilde{H}_{\rm{S}}$, it is clear that only the bright state, the single-exciton state with zero momentum $|k=0\rangle=a^\dag_0|\rm{vac}\rangle$, couples to the cavity photon. The $N-1$ dark states $|k\rangle=a^\dag_k|\rm{vac}\rangle$ ($k\neq0$) are themselves eigenstates of $\tilde{H}_{\rm{S}}$. The interaction between the bright exciton and the cavity mode results in two eigenmodes which bring $\tilde{H}_{\rm{S}}$ into a diagonal form,
\begin{eqnarray}\label{Hsud}
\tilde{H}_{\rm{S}}&=&\sum_{k\neq0}E_k a^\dag_ka_k+E_{\rm{U}} a_{\rm{U}}^\dag a_{\rm{U}}+E_{\rm{D}} a_{\rm{D}}^\dag a_{\rm{D}},
\end{eqnarray}
where $a_{\rm{U}}^\dag=Ca^\dag_0-Sc^\dag$ and $a_{\rm{D}}^\dag=Sa^\dag_0+Cc^\dag$ are the creation operators of two new quasiparticles, and the corresponding eigenenergies are
\begin{eqnarray}\label{Eud}
E_{\rm{U/D}}=\frac{E_0+\tilde{\omega}_{\rm{c}}}{2}\pm\sqrt{N\tilde{g}^2+\left(\frac{E_0-\tilde{\omega}_{\rm{c}}}{2}\right)^2}.
\end{eqnarray}
Here, the mixing coefficients $C=\cos\frac{\theta}{2}$ and $S=\sin\frac{\theta}{2}$ are determined by
\begin{eqnarray}\label{Tan}
\tan\theta=2\tilde{g}\sqrt{N}/(\tilde{\omega}_{\rm{c}}-E_0).
\end{eqnarray}
Although the two branches of eigenmodes resemble the lower and upper exciton polaritons~\cite{LP1992}, we have to keep in mind that these structures appear in the Merrifield frame, and hence do not correspond to physical quasiexcitations. Actually, transforming back to the original frame from the Merrifield frame will yield \emph{physical} quasiparticles which are mixtures of excitonic, photonic, and vibrational degrees of freedom. In the following, we will refer to the $U$ and $D$ quasiparticles as \emph{Merrifield polaritons}.
\par To obtain the optimal zero-order Hamiltonian $\tilde{H}_0=\tilde{H}_{\rm{S}}+H_{\rm{v}}$, we proceed by minimizing the Bogoliubov upper bound for the free energy of $\tilde{H}$ at inverse temperature $\beta$,
\begin{eqnarray}\label{BB}
F_{\rm{B}}=-\frac{1}{\beta}\ln \rm{Tr} e^{-\beta\tilde{H}_0}+\langle\tilde{V}\rangle_0,
\end{eqnarray}
where $\langle...\rangle_0=\rm{Tr}\{...e^{-\beta\tilde{H}_0}\}/Tr\{e^{-\beta\tilde{H}_0}\}$. By construction, $\langle\tilde{V}\rangle_{\rm{0}}=0$, so that the second term in Eq.~(\ref{BB}) vanishes. In the single-excitation subspace, the Bogoliubov bound can be expressed in terms of single-particle eigenenergies of $\tilde{H}_0$ as
\begin{eqnarray}\label{BBs}
F_{\rm{B}}=-\frac{1}{\beta}\ln Z_{\rm{S}}+F_{\rm{v}},
\end{eqnarray}
where
\begin{eqnarray}\label{PF}
Z_{\rm{S}}=\sum_{\eta=\{ k(\neq0),\rm{U,D}\}} e^{-\beta E_\eta},
\end{eqnarray}
is the partition function for $\tilde{H}_{\rm{S}}$, and $F_{\rm{v}}=-\frac{1}{\beta}\ln Z_{\rm{v}}$ is the free energy of the free vibrational modes. As $F_{\rm{v}}$ is not dependent on the variational parameters, we only need to minimize the first term of Eq.~(\ref{BBs}). To this end, the saddle-point conditions $\{\partial F_{\rm{B}}/\partial f_n=0\}$ and $\partial F_{\rm{B}}/\partial h=0$ should be solved self-consistently. Two forms of the resultant saddle-point equations are listed in Appendix \ref{AppA}. We emphasize that the such obtained $F_{\rm{B}}$ gives an upper bound for the intrinsic free energy of the system.
\par It is convenient to write the residue interaction $\tilde{V}$ in the basis $\{|\eta\rangle=a^\dag_\eta|\rm{vac}\rangle\}$ ($\eta=k(\neq0),\rm{U},\rm{D}$),
\begin{eqnarray}\label{Vetaeta}
\tilde{V}&=&\sum_{\eta_1\eta_2}|\eta_1\rangle\langle\eta_2|\tilde{V}_{\eta_1\eta_2},\nonumber\\
\tilde{V}_{\eta_1\eta_2}&=&x_{\eta_1}x_{\eta_2}[P_{k(\eta_1),k(\eta_2)}+T_{k(\eta_1)-k(\eta_2)}]\nonumber\\
&&+x_{\eta_1}y_{\eta_2}R_{k(\eta_1)}+y_{\eta_1}x_{\eta_2}R^\dag_{k(\eta_2)}\nonumber\\
&&+y_{\eta_1}y_{\eta_2}\omega_0h\sqrt{N}(b_0+b^\dag_0),
\end{eqnarray}
where
\begin{eqnarray}\label{xy_ind}
\{x_\eta\}=\{1,...,1,C,S\},~\{y_\eta\}=\{0,...,0,-S,C\},
\end{eqnarray}
and $\{k(\eta)\}=\{k(\neq0),0,0\}$.
\section{Ground state: The lower polaron-polariton}\label{III}
\par For the Holstein model without the cavity, one can introduce the adiabaticity ratio $\gamma=\omega_0/|J|$ and the dimensionless exciton-vibration coupling strength $\alpha=\frac{1}{2}\gamma\lambda^2$. Then, $\gamma<1$ ($>1$) defines the adiabatic (antiadiabatic) regime, and $\alpha >1$ ($<1$) defines the strong (weak) exciton-vibration coupling regime~\cite{PRB_hol_dim}. Besides the method employed in the present work, the ground state of the Holstein model has been widely studied by various analytical/numerical methods, including numerical diagonalization based on the two-particle approximation~\cite{Philpott,Soos,Spano2002,Spano2014}, quantum Monte Carlo simulation~\cite{PRB1984}, density matrix renormalization-group technique~\cite{PRB1998}, exact-diagonalization method~\cite{PRB1994, PRB1997}, and variational ansatz~\cite{JCP1997}.

\par In the presence of the cavity and in the zero-temperature limit, the Bogoliubov bound to be minimized becomes the zero-order ground-state energy $E_{\rm{D}}$, i.e., the eigenenergy of the lower Merrifield polariton,
\begin{eqnarray}\label{LP_MP}
|\rm{D}\rangle=a^\dag_{\rm{D}}|\rm{vac}\rangle.
\end{eqnarray}
\par In this case, it can be shown that the optimal variational parameters are given by (see Appendix \ref{AppA})
\begin{eqnarray}\label{lb_h_0k}
\frac{\lambda}{Nh}&=&\frac{1}{S^2}-\frac{\omega_0}{\tilde{g}\sqrt{N}}\frac{C}{S},
\end{eqnarray}
\begin{eqnarray}\label{lb_f0_0k}
\frac{\tilde{f}_0}{Nh}=1-\frac{\omega_0}{\tilde{g}\sqrt{N}}\frac{C}{S},
\end{eqnarray}
and
\begin{eqnarray}\label{lb_fq_0k}
\frac{\lambda}{\tilde{f}_q}
&=&1-\frac{\tilde{g}\sqrt{N}}{\omega_0}\frac{C}{S}-2(1-\cos q)\frac{\tilde{J}}{\omega_0},
\end{eqnarray}
for $q\neq0$.

\par In the absence of the vibrational modes, the ground state of the exciton-cavity system is simply obtained by diagonalizing the Hamiltonian $H_{\rm{e}}+H_{\rm{c}}+H_{\rm{e-c}}$, yielding the bare upper polariton and lower polariton,
\begin{eqnarray}\label{LP_bare}
|\phi_{\rm{UP}}\rangle=C_0|0\rangle-S_0|c\rangle,~|\phi_{\rm{LP}}\rangle=S_0|0\rangle+C_0|c\rangle,
\end{eqnarray}
where $C_0=\cos\frac{\theta_0}{2},~S_0=\sin\frac{\theta_0}{2}$ with $\tan\theta_0=2g\sqrt{N}/(\omega_{\rm{c}}-\varepsilon_0-2J)$. When the vibrational bath is present, the corresponding physical ground state can be obtained by transforming $|\rm{D}\rangle$ back to the original frame,
\begin{eqnarray}\label{LP}
|\psi_{\rm{LPP}}\rangle&=&e^{-\mathcal{S}}|\rm{D}\rangle\nonumber\\
&=& \frac{S}{N}\sum_n\sum_k e^{-ikn} e^{-\frac{1}{\sqrt{N}}\sum_q(\tilde{f}_{q}e^{-iqn}b^\dag_q-\tilde{f}_{-q}e^{iqn}b_q)}|k\rangle\nonumber\\
&&+C e^{-h\sqrt{N}  (b^\dag_0-b_0)} |c\rangle.
\end{eqnarray}
We see that $|\psi_{\rm{LPP}}\rangle$ has a similar structure to the free LP state given by Eq.~(\ref{LP_bare}), but includes the vibration-induced effects through the vibrational coherent states. The first term on the right-hand side of Eq.~(\ref{LP}) mimics a polaron state with amplitude $S$, while the second term corresponds to the vibrational dressing of the cavity state with amplitude $C$. Furthermore, unlike the bare LP state which only has a component of the bright exciton $|0\rangle$, the exciton-vibration coupling also mixes the excitonic dark states $|k\rangle$ ($k\neq0$) into $|\psi_{\rm{LPP}}\rangle$. For this reason, we refer to the quasiparticle corresponding to $|\psi_{\rm{LPP}}\rangle$ as the \emph{lower polaron polariton} (LPP).

\par The above form of the LPP state can be compared with the following generalized Toyozawa ansatz~\cite{PRB_hol_dim,Toyozawa1961}:
\begin{eqnarray}\label{TA}
|\psi_{\rm{TA}}\rangle&=&\frac{1}{\sqrt{N}}\sum_n\sum_k e^{-ikn}\Phi_k e^{\sum_q(\xi_q e^{-iqn}b^\dag_{q}-\xi^{*}_q e^{iqn}b_q)}|k\rangle\nonumber\\
&&+\frac{1}{\sqrt{N}}\Phi_{\rm{c}} e^{ \xi_{\rm{c}} b^\dag_{0}-\xi_{\rm{c}}^{*} b_0 }|c\rangle,
\end{eqnarray}
which recovers $|\psi_{\rm{LPP}}\rangle$ in the case of
\begin{eqnarray}
\Phi_k&=&\frac{S}{\sqrt{N}},~\xi_q=-\frac{\tilde{f}_q}{\sqrt{N}},\nonumber\\
\Phi_{\rm{c}}&=&C\sqrt{N},~\xi_{\rm{c}}=-h\sqrt{N}.
\end{eqnarray}

\par The Toyozawa ansatz (TA) is believed to provide accurate results for the ground-state wave function and ground-state energy of the Holstein model~\cite{PRB_hol_dim,JCP1997,TA2012}. Since there are more variational parameters in $|\psi_{\rm{TA}}\rangle$ than those in $|\psi_{\rm{LPP}}\rangle$, the ground-state energy $E_{\rm{TA}}$ obtained by TA is slightly lower than $E_{\rm{D}}$. However, we emphasize that the Merrifield transformation based on minimizing the Bogoliubov free energy also applies to \emph{finite temperatures}.
\begin{figure}
(a)\includegraphics[width=.50\textwidth]{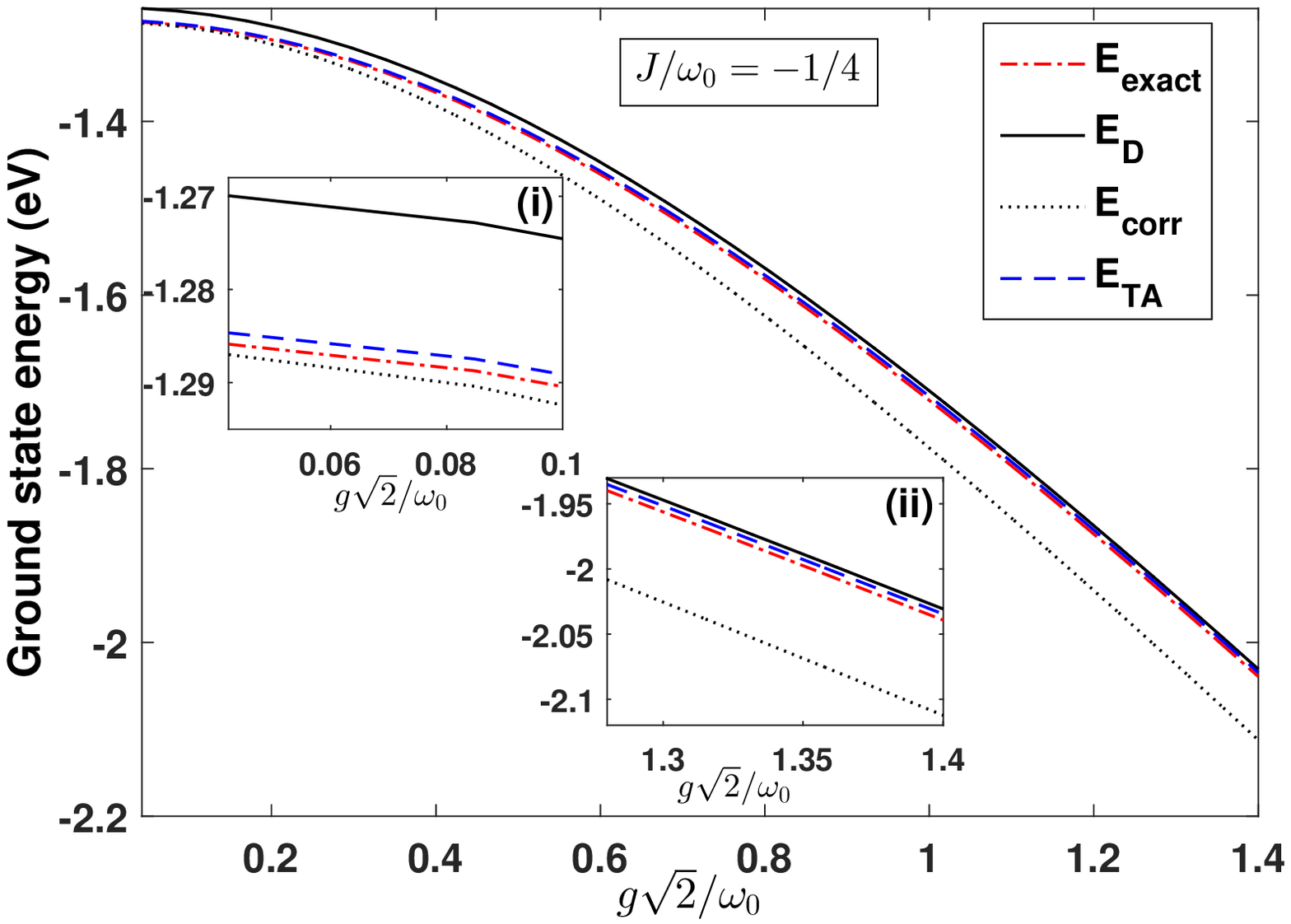}
(b)\includegraphics[width=.50\textwidth]{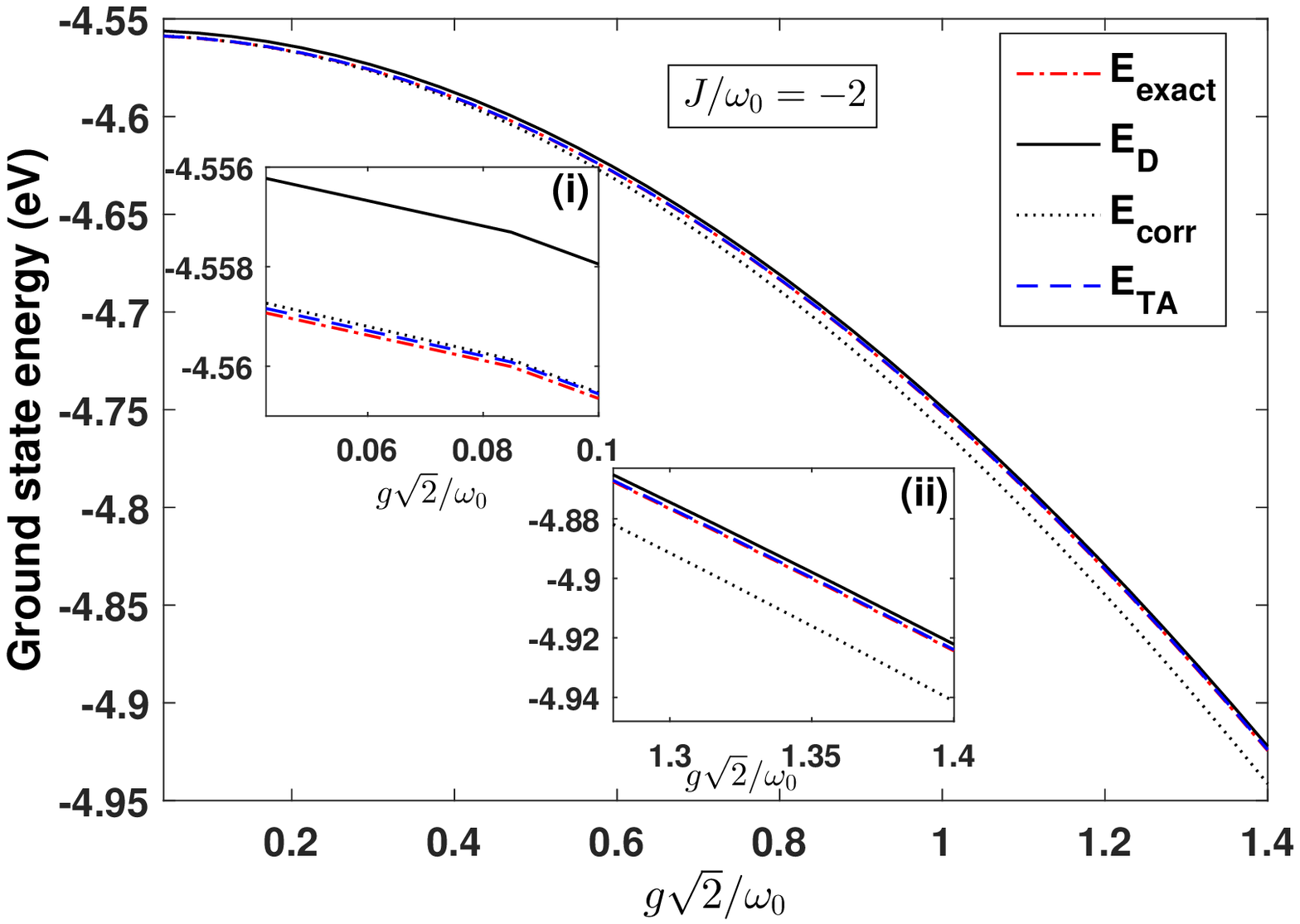}
\caption{The ground-state energy of a molecular dimer interacting with a single cavity mode as a function of the collective exciton-cavity coupling $g\sqrt{N}/\omega_0$ for two different sets of excitonic coupling: (a) $J/\omega_0=-1/4$ and (b) $J/\omega_0=-2$. Results from the exact numerical diagonalization (red line), the variational Merrifield transformation without and with the second-order energy correction (solid black and dotted black lines), and the Toyozawa ansatz (blue line) are presented. The insets in each figure show the magnifications in the (i) weak and (ii) strong exciton-cavity regimes. Other parameters: $\lambda=1$, $\omega_0=1 eV$, $\varepsilon_0=\omega_{\rm{c}}=0 eV$, and $M_{\max}=20$.}
\label{dimer}
\end{figure}

\par From the variational principle, $E_{\rm{D}}$ provides an upper bound for the true ground-state energy of the system. A conventional procedure for obtaining a lower approximated ground-state energy is to calculate the second-order energy correction in terms of the residue interaction $\tilde{V}$~\cite{Cheng2008}. For small systems, we have checked numerically that the second-order correction to $E_{\rm{D}}$ gives a more accurate approximation for $g\sqrt{N}/\omega_0\ll1$, but underestimates the true ground-state energy in the strong exciton-cavity coupling regime with relatively large $g\sqrt{N}/\omega_0$. This can be illustrated by studying a molecular dimer with $N=2$ chromophores, for which the ground-state energy can be obtained exactly by numerically diagonalizing the Hamiltonian in a truncated vibrational space with $\sum_{i=1,2}b^\dag_i b_i=M_{\max}$ vibrations.

\par  Figure \ref{dimer} shows the calculated ground-state energy of a molecule dimer by the exact numerical diagonalization with up to $M_{\max}=20$ vibrations ($E_{\rm{exact}}$), the variational Merrifield transformation without and with the second-order energy correction ($E_{\rm{D}}$ and $E_{\rm{corr}}$; see Appendix \ref{2nd}), as well as the Toyozawa ansatz ($E_{\rm{TA}}$). We set $\lambda=1$, $\omega_0=1 eV$, and $\varepsilon_0=\omega_{\rm{c}}=0 eV$, namely, a cavity frequency resonant with the on-site excitonic transition. The results for two sets of nearest-neighbor interactions $J/\omega_0=-1/4$ and $J/\omega_0=-2$ are presented in Fig.~\ref{dimer}(a) and Fig.~\ref{dimer}(b), which correspond to the antiadiabatic strong exciton-vibration coupling limit and adiabatic weak-coupling limit, respectively.

\par Insets (i) and (ii) in Fig.~\ref{dimer}(a) and (b) display the magnification in the weak and strong exciton-cavity coupling region, respectively. In both cases, we find that the zero-order energy $E_{\rm{D}}$ from the Merrifield transformation overestimates the ground state energy in the weak exciton-cavity regime $g\sqrt{2}/\omega_0\ll1$, while the second-order corrected energy $E_{\rm{corr}}$ gives a more accurate one. However, $E_{\rm{corr}}$ begins to show large deviation from the exact value $E_{\rm{exact}}$ and underestimates the true ground-state energy when one enters the strong cavity coupling regime $g\sqrt{2}/\omega_0\sim 1$. In contrast, both the zeroth-order energy $E_{\rm{D}}$ and the Toyozawa variational energy $E_{\rm{TA}}$ become closer to $E_{\rm{exact}}$ in  this regime. Since the first-order energy correction vanishes by construction, while the second-order energy correction is always negative, it is expected that higher order corrections are needed to get a more accurate ground-state energy for large $g\sqrt{N}/\omega_0$. As we are mainly interested in the strong exciton-cavity coupling regime, we will henceforth take
\begin{eqnarray}
E_{\rm{LPP}}\approx E_{\rm{D}}
\end{eqnarray}
as an approximation of the ground-state energy. Correspondingly, we take the LPP wave function given by Eq.~(\ref{LP}) as an approximated ground state in order to obtain simple and intuitive analytical expressions for observables discussed in the following.

\par
\begin{figure}
(a)\includegraphics[width=.50\textwidth]{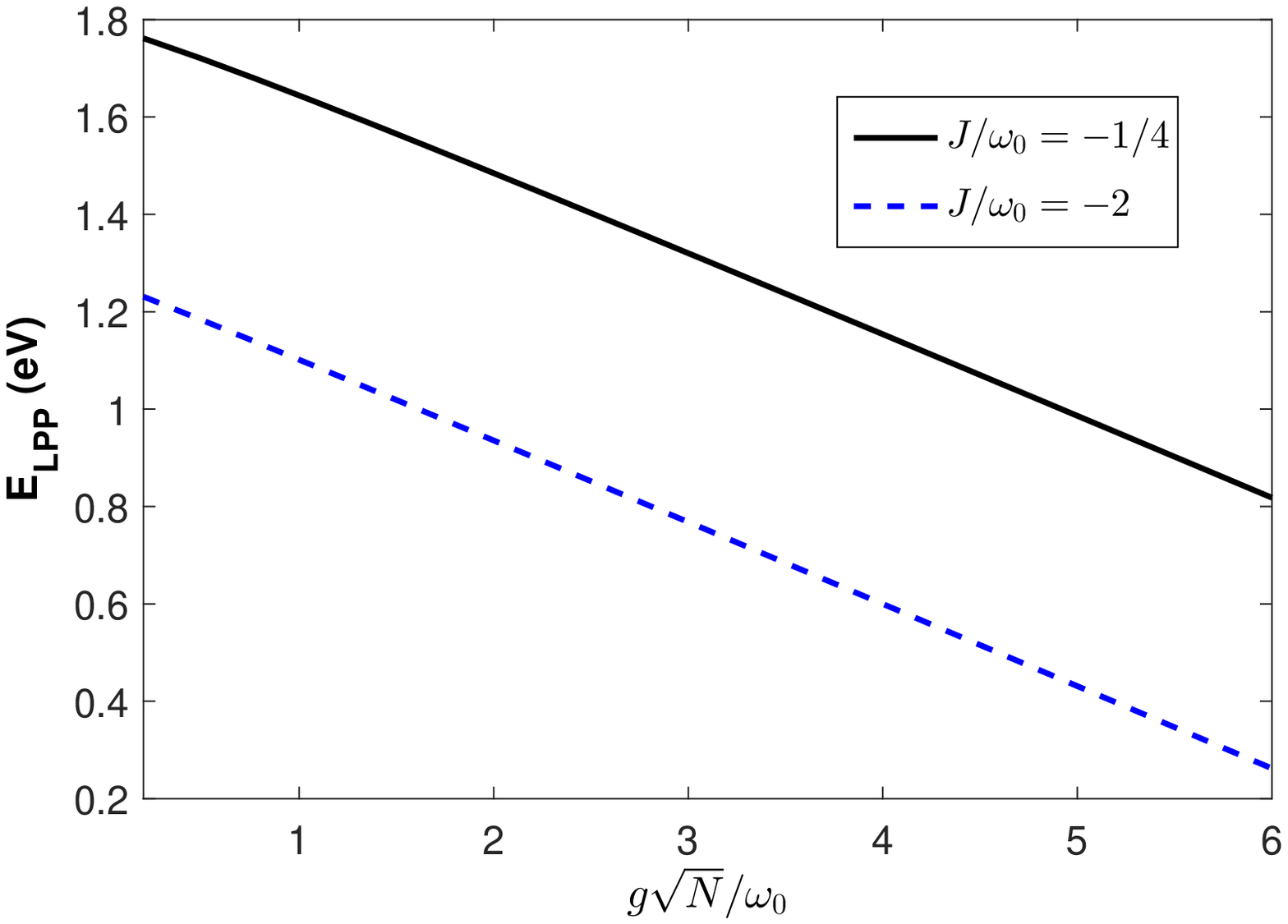}
(b)\includegraphics[width=.50\textwidth]{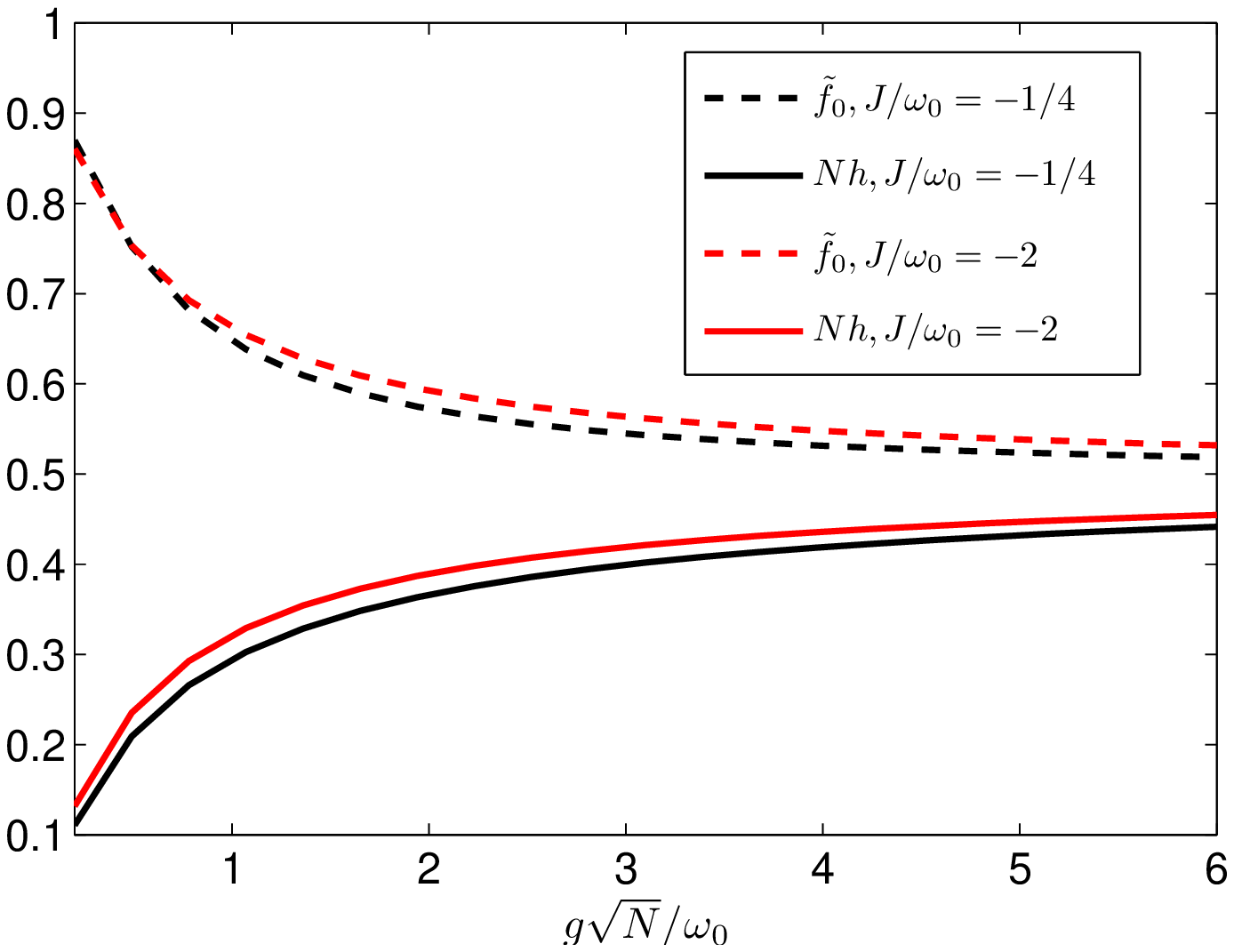}
\caption{(a) The ground-state energy $E_{\rm{LPP}}$ of the LPP as a function of the collective exciton-cavity coupling $g\sqrt{N}/\omega_0$ for two different excitonic couplings: $J/\omega_0=-1/4$ and $J/\omega_0=-2$. (b)  The evolution of the collective vibrational dressing parameter $\tilde{f}_0$ of the excitons and the cavity dressing parameter $Nh$ with $g\sqrt{N}/\omega_0$. Other parameters: $N=100$, $\omega_0=0.17 eV$, $\varepsilon_0=2 eV$, and $\lambda=1$. In both cases, the cavity frequency is set to be resonant with the ground-state energy of the molecule in the absence of the cavity.}
\label{FIG1}
\end{figure}

  Figure~\ref{FIG1}(a) shows the ground-state energy $E_{\rm{LPP}}$ of the LPP as a function of the dimensionless exciton-cavity coupling strength $g\sqrt{N}/\omega_0$ for two different excitonic couplings $J/\omega_0=-1/4$ and $J/\omega_0=-2$. Other molecular parameters are taken as $N=100$, $\omega_0=0.17 eV$, $\varepsilon_0=2 eV$, and $\lambda=1$. Note that $J/\omega_0=-1/4$ is a typical value for J-aggregates~\cite{Spano2015} and the molecular system thus lies in the \emph{small-polaron} limit, i.e., the strong exciton-vibration coupling antiadiabatic limit~\cite{PRB_hol_dim}. The cavity frequency is set to be resonant with the ground state energy in the absence of the cavity [which is calculated by using the Toyozawa ansatz given by Eq.~(\ref{TA}) in the $g\to0$ limit for $N=100$], i.e., $\omega_{\rm{c}}=E_{\rm{TA}}(N=100,g=0)=1.7864 eV$ and $1.2617eV$ for $J/\omega_0=-1/4$ and $J/\omega_0=-2$, respectively. As expected, the coupling between the bright exciton and the cavity mode leads to the formation of the LPP state which lies below the pure polaron state of the molecule.

\begin{figure}
\includegraphics[width=.65\textwidth]{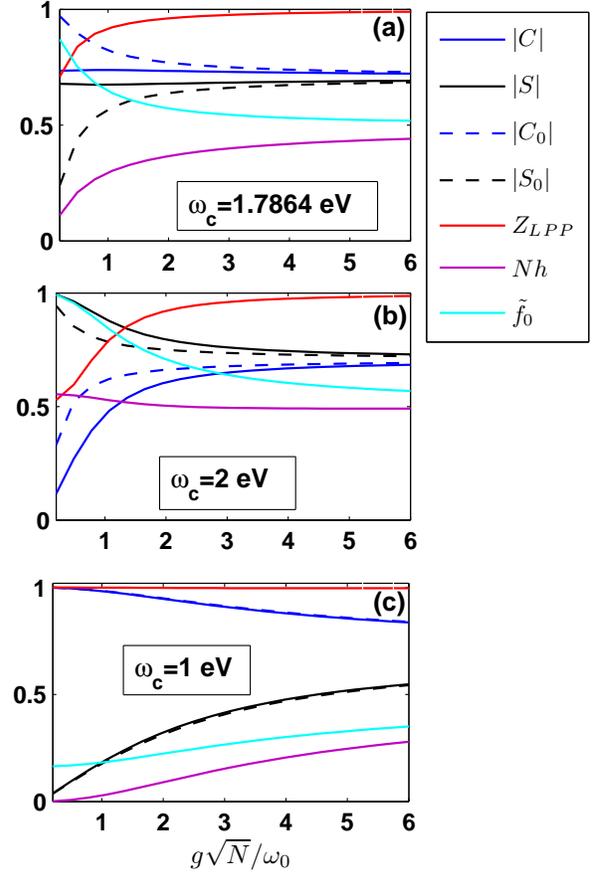}
\caption{Absolute values of the mixing coefficients $C$ and $S$ ($C_0$ and $S_0$) in the LPP state $|\psi_{\rm{LPP}}\rangle$ (the bare LP state $|\phi_{\rm{LP}}\rangle$), and the quasiparticle weight $Z_{\rm{LPP}}$ as functions of $g\sqrt{N}/\omega_0$ for (a) a resonant cavity mode with $\omega_{\rm{c}}=1.7864eV$, (b) $\omega_{\rm{c}}=2eV$, (c) $\omega_{\rm{c}}=1eV$. Also shown are the dressing parameters $Nh$ and $\tilde{f}_0$. Other parameters: $N=100$, $\omega_0=0.17 eV$, $\varepsilon_0=2 eV$, $J/\omega_0=-1/4$, and $\lambda=1$.}
\label{FIG2}
\end{figure}

\par In Fig.~\ref{FIG1}(b), we plot the evolution of $\tilde{f}_0=\sum_n f_n$ and the cavity dressing parameter $Nh$ with the exciton-cavity coupling $g\sqrt{N}/\omega_0$. The decreasing of $\tilde{f}_0$ with increasing $g$ clearly indicates a reduced vibrational dressing of excitons in the strong-coupling regime. We note that the decoupling of vibrational degrees of freedom from the excitons by strong cavity coupling has also been reported by Spano and co-workers~\cite{Spano2015,Spano2016} by using numerical diagonalization of the Holstein Hamiltonian. It is intriguing to note that for this \emph{resonant} case, the cavity dressing parameter $Nh$ increases monotonically as $g$ increases, which means that the cavity mode, even though it is \emph{not} coupled to the vibrations \emph{directly}, might become more dressed by the vibrations as the exciton-cavity coupling increases.

\par However, as can be seen from Eq.~(\ref{LP}), the degree of the vibrational dressing of the cavity is measured by both the amplitude $C$ and the dressing parameter $Nh$. To this end, we plot in Fig.~\ref{FIG2} the absolute values of the amplitudes $C$ and $S$ ($C_0$ and $S_0$) in the LPP $|\psi_{\rm{LPP}}\rangle$ (the bare LP $|\phi_{\rm{LP}}\rangle$) as functions of $g\sqrt{N}/\omega_0$ for both resonant and nonresonant cases. The behavior in the weak cavity coupling region $g\sqrt{N}/\omega_0\ll1$ can be understood from investigating the saddle-point equations (\ref{lb_h_0k}) and (\ref{lb_f0_0k}). As $g\sqrt{N}/\omega_0\to0^+$, we have
\begin{eqnarray}\label{Nh_g0}
Nh\approx\lambda\frac{\theta(\tilde{\omega}_{\rm{c}}-E_0)}{1+\omega_0/[2(\tilde{\omega}_{\rm{c}} -E_0)]},
\end{eqnarray}
and
\begin{eqnarray}\label{f0_g0}
\tilde{f}_0\approx \frac{\lambda}{1+2(E_0-\tilde{\omega}_{\rm{c}})\theta(E_0-\tilde{\omega}_{\rm{c}})/\omega_0},
\end{eqnarray}
where $\theta(x)$ is the Heaviside step function.

\par For the resonant case, the polaron part and the dressed cavity part of $|\psi_{\rm{LPP}}\rangle$ are roughly equally weighted for weak exciton-cavity coupling [Fig.~\ref{FIG2}(a)] with $|C|\approx|S|\approx1/\sqrt{2}$. As $g$ increases, the cavity dressing parameter $Nh$ increases monotonically, while the amplitude $|C|$ has no dramatic change even up to the strong-coupling regime. For the non-resonant case with the cavity frequency $\omega_{\rm{c}}$ relatively large, so that the condition $\tilde{\omega}_{\rm{c}}-E_0>0$ is fulfilled [Fig.~\ref{FIG2}(b)], we observe a slow drop of $Nh$ with increasing $g$. However, the increase of the amplitude $|C|$ from $0^+$ to its saturated value in the strong-coupling region might still indicate an enhanced dressing of the cavity field. For the nonresonant case with a red-detuned cavity frequency, both the collective molecular dressing parameter $\tilde{f}_0$ and the cavity dressing parameter increase as $g\sqrt{N}/\omega_0$ increases, so it is expected that the exciton-cavity coupling can enhance the dressing of both the exciton and the cavity. As we will see in the next section, a better measure for the degree of vibrational dressing is the mean vibration number on a specific exciton/cavity state, which involves both the amplitude and the dressing parameter.
\begin{figure}
\includegraphics[width=.55\textwidth]{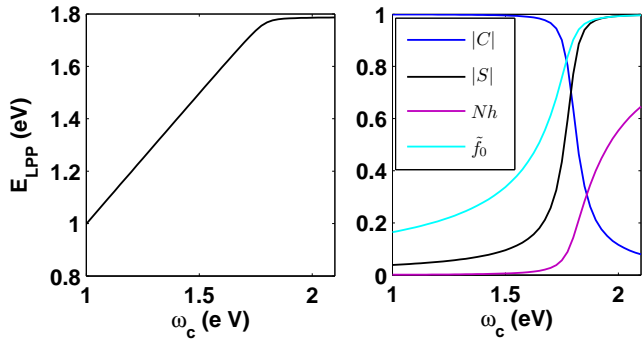}
\caption{Left panel: The ground-state energy $E_{\rm{LPP}}$ as a function of the cavity frequency $\omega_{\rm{c}}$. Right panel: Absolute values of the amplitudes $|C|$ and $|S|$, as well as the parameters $\tilde{f}_0$ and $Nh$ as functions of $\omega_{\rm{c}}$. The exciton-cavity coupling is set to be in the weak-coupling regime as $g\sqrt{ N}/\omega_0=0.2$. Other parameters: $N=100$, $\omega_0=0.17 eV$, $\varepsilon_0=2 eV$, $J/\omega_0=-1/4$, and $\lambda=1$.}
\label{FIGwc}
\end{figure}

\par In order to see the crossover with the cavity detuning in the weak cavity coupling region more clearly, we plot in the left panel of Fig.~\ref{FIGwc} the ground-state energy as a function of the cavity frequency $\omega_{\rm{c}}$ for fixed exciton-cavity coupling strength $g\sqrt{ N}/\omega_0=0.2$. A crossover can be seen around the resonant frequency $1.7864eV$. The right panel of Fig.~\ref{FIGwc} shows the behavior of the amplitudes $|C|$ and $|S|$, and the dressing parameters $Nh$ and $\tilde{f}_0$. For a red-detuned cavity with frequency $\omega_{\rm{c}}<1.7864eV$, the LPP state $|\psi_{\rm{LPP}}\rangle$ is dominated by the cavity component with a small vibrational dressing. For a blue-detuned cavity, the ground state behaves more like a polaron. However, the cavity becomes more dressed though its amplitude $|C|$ decreases with increasing $\omega_{\rm{c}}$. Note that the profiles of $Nh$ and $\tilde{f}_0$ approach the limiting forms given by Eqs.~(\ref{Nh_g0}) and (\ref{f0_g0}) if $g\sqrt{N}/\omega_0$ is lowered down further.

\par Intriguing enough, it can be seen from Fig.~\ref{FIG2} that $C$ and $S$ tend to be consistent with $C_0$ and $S_0$ as $g\sqrt{N}/\omega_0$ increases for all the three cases considered. This behavior can be better understood by introducing the following quasiparticle weight for the LPP:
\begin{eqnarray}
Z_{\rm{LPP}}=|\langle\phi_{\rm{LP}}|\psi_{\rm{LPP}}\rangle|^2,
\end{eqnarray}
which measures how similar the LPP wavefunction $|\psi_{\rm{LPP}}\rangle$ is to the vibration-free LP wave function $|\phi_{\rm{LP}}\rangle$. It is easy to show that
\begin{eqnarray}
Z_{\rm{LPP}}&=&|S_0S\Theta_0+C_0C e^{-\frac{1}{2}h^2N}|^2,
\end{eqnarray}
where $\Theta_0=\Theta(h=0)$. As can be seen in Fig.~\ref{FIG2}, $Z_{\rm{LPP}}$ approaches nearly unity monotonically as the exciton-cavity coupling increases, which means that the LPP behaves like a vibration-free LP in the strong-coupling regime.

\par Actually, by investigating Eqs.~(\ref{lb_h_0k})--(\ref{lb_fq_0k}) in the ultrastrong-coupling limit $g\sqrt{N}/\omega_0\to\infty$, we have  $C\approx-S\approx1/\sqrt{2}$, $Nh\approx\tilde{f}_0\approx\lambda S^2\approx\lambda/2$, and $\tilde{f}_q\approx0$ ($q\neq0$), which gives the asymptotic form of $|\psi_{\rm{LPP}}\rangle$,
\begin{eqnarray}\label{LP_ginf}
|\psi_{\rm{LPP}}\rangle&\approx&\frac{1}{\sqrt{2}}(|c\rangle-|0\rangle)e^{-\frac{\lambda}{2\sqrt{N}} (b^\dag_0-b_0)}|\rm{vac_{v}}\rangle\nonumber\\
&\approx&|\phi_{\rm{LP}}\rangle e^{-\frac{\lambda}{2\sqrt{N}} (b^\dag_0-b_0)}|\rm{vac_{v}}\rangle,
\end{eqnarray}
where $|\rm{vac_{v}}\rangle$ denotes the vibrational vacuum state.
\par Equation~(\ref{LP_ginf}) indicates that in the ultrastrong-coupling limit, the LPP state tends to be a \emph{separable state}, which is consistent with the bare LP state dressed by the zero-momentum vibrational mode. Furthermore, the vibrational dressing part becomes negligible for large aggregates with $N\gg1$. Correspondingly, the quasiparticle weight approaches
\begin{eqnarray}\label{ZLPP_strog}
Z_{\rm{LPP}}&\approx&e^{-\frac{\lambda^2}{4N}},~g\sqrt{N}/\omega_0\to\infty.
\end{eqnarray}
\section{The photoluminescence line strength and the mean number of vibrations}\label{IV}
\subsection{Zero temperature}
\par In order to better understand the influence of strong exciton-cavity coupling on the molecular system, it is instructive to study the variation of several observables with the cavity coupling strength. In this section, we calculate the photoluminescence line strength and the mean number of vibrations using the results obtained in the last section.

\par After photoexcitation, a J-aggregate loses its excess energy and reaches the bottom of the exciton band quickly, so that the emission process originates mainly near the band bottom. When the cavity mode is present, the LPP state $|\psi_{\rm{LPP}}\rangle$ takes the role of such a band bottom exciton. The $0-\xi$ photoluminescence line strength $I^{0-\xi}$ arising from transitions between $|\psi_{\rm{LPP}}\rangle$ and the excitonic ground state with $\xi$ vibrations is defined as~\cite{Spano2015}
\begin{eqnarray}\label{PL_xi}
I^{0-\xi}&=&\frac{1}{\mu^2}\sum_{\sum_{q}n_q=\xi}| \langle \{n_q\}|\hat{\mu}|\psi_{\rm{LPP}}\rangle|^2,
\end{eqnarray}
where the transition dipole moment operator is given by
\begin{eqnarray}\label{mu}
\hat{\mu}&=&\mu\sum_n(|n\rangle\langle \rm{vac}|+|\rm{vac}\rangle\langle n|).
\end{eqnarray}
By inserting Eqs.~(\ref{LP}) and (\ref{mu}) into Eq.~(\ref{PL_xi}), we obtain (see Appendix \ref{AppB})
\begin{eqnarray}\label{Ixi}
I^{0-\xi}&=&\frac{(S\Theta_0)^2 }{\xi!N^{\xi}}\sum_n(G_n)^\xi,
\end{eqnarray}
where
\begin{eqnarray}\label{Gn}
G_n=G^*_n=\sum_q e^{iqn} \tilde{f}_{q}\tilde{f}_{-q}.
\end{eqnarray}
The first three cases for $\xi=0,1$, and $2$ can be calculated as
\begin{eqnarray}
I^{0-0}&=&N(S\Theta_0)^2,\\
I^{0-1}&=&(S\Theta_0\tilde{f}_0)^2,\\
I^{0-2}&=&\frac{(S\Theta_0)^2 }{2N} \sum_{q}  (\tilde{f}_{q}\tilde{f}_{-q})^2.
\end{eqnarray}
\begin{figure}
\includegraphics[width=.55\textwidth]{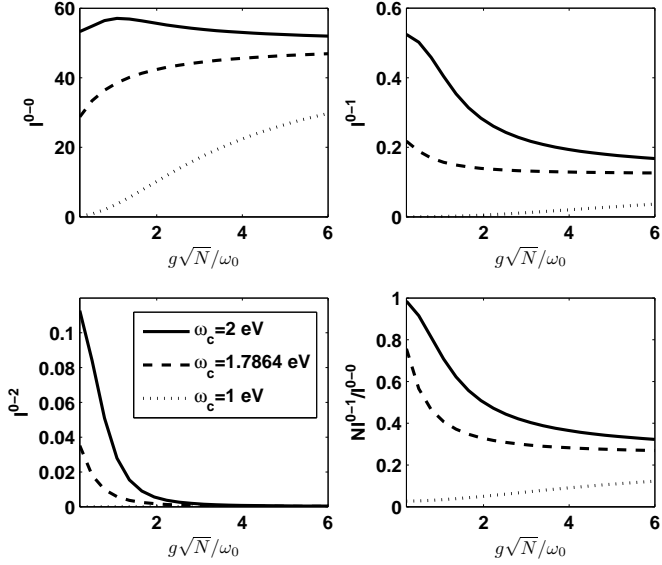}
  \caption{The zero-temperature photoluminescence line strength $I^{0-0}$, $I^{0-1}$, $I^{0-2}$ and the strength ratio $N I^{0-1}/I^{0-0}$ calculated using Eq.~(\ref{Ixi}). The results for three sets of cavity frequencies $\omega_{\rm{c}}=2 eV$ (solid line), $\omega_{\rm{c}}=1.7864 eV$ (dashed line), and $\omega_{\rm{c}}=1 eV$ (dotted line) are shown. Other parameters: $N=100$, $\omega_0=0.17 eV$, $\varepsilon_0=2 eV$, $J/\omega_0=-1/4$, and $\lambda=1$.}
\label{FIG3}
\end{figure}

\par The zero-temperature $0-\xi$ photoluminescence line strength $I^{0-\xi}$ for $\xi=0,1$, and $2$ is shown in the first three panels of Fig.~\ref{FIG3}, where results for both the resonant and nonresonant cases are presented. For the blue-detuned cavity frequency $\omega_{\rm{c}}=2eV$, we observe a nonmonotonic behavior in the $0-0$ line strength $I^{0-0}$ with increasing $g$. For the resonant and red-detuned cavity, an amplification of $I^{0-0}$ is observed. The last panel of Fig.~\ref{FIG3} shows the line strength ratio,
\begin{eqnarray}
N \frac{I^{0-1}}{I^{0-0}}&=&\tilde{f}_0^2,
\end{eqnarray}
which is proportional to an effective Huang-Rhys factor~\cite{Spano2015} and decreases with increasing exciton-cavity coupling for resonant and blue-detuned cavity frequencies. However, the increase of this ratio for a red-detuned cavity shows that the cavity coupling can actually increase the effective exciton-vibration coupling from weak to intermediate cavity coupling regions.

\par The above results can be further demonstrated by studying the mean number of vibrations in the LPP state, which is an important measure of vibrational dressing of specific exciton-cavity states~\cite{Silbey19771},
\begin{eqnarray}\label{Nv}
N_{\rm{v}}&=&\langle \psi_{\rm{LPP}}|\sum_{q}b^\dag_q b_q|\psi_{\rm{LPP}}\rangle.
\end{eqnarray}
Straightforward calculation gives (see Appendix \ref{AppB})
\begin{eqnarray}\label{Nv_re}
N_{\rm{v}}&=&(S\Theta_0)^2\frac{G_0}{N}e^{\frac{G_0}{N}}+N C^2 h^2\nonumber\\
&=&NN^{(\rm{site})}_{\rm{v}}+N^{(\rm{c})}_{\rm{v}}
\end{eqnarray}
where
\begin{eqnarray}\label{Npha}
N^{(\rm{c})}_{\rm{v}}&=&N C^2h^2,
\end{eqnarray}
is the mean number of vibrations projected onto the cavity state $|c\rangle$, and
\begin{eqnarray}\label{Nvsite}
N^{(\rm{site})}_{\rm{v}}&=&(S\Theta_0)^2\frac{G_0}{N^2}e^{\frac{G_0}{N}}
\end{eqnarray}
is the mean number of vibrations in the cloud surrounding a local exciton, which is identical for all sites due to the circular symmetry of the molecule.
\begin{figure}
(a)\includegraphics[width=.53\textwidth]{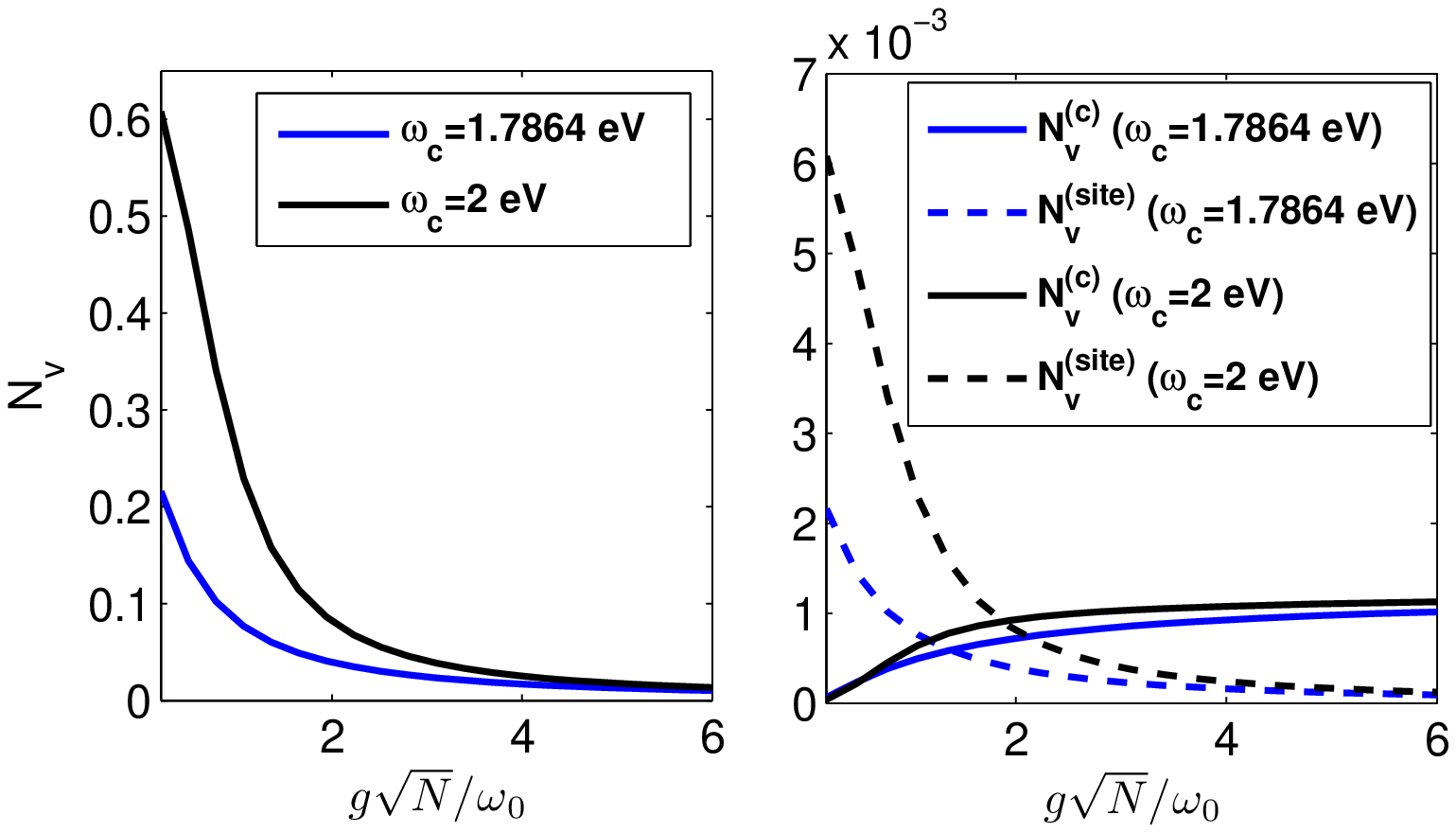}
(b)\includegraphics[width=.53\textwidth]{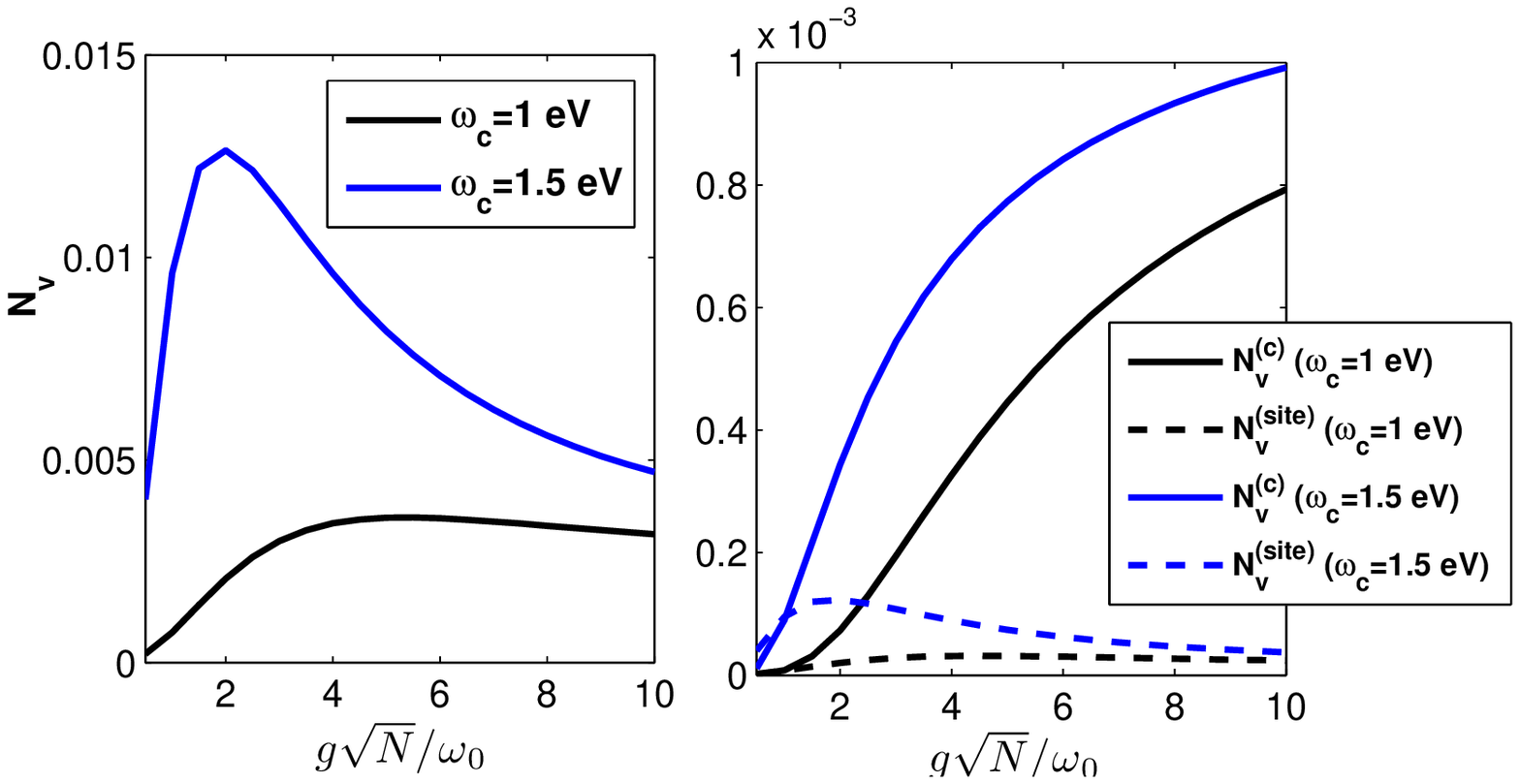}
  \caption{The total mean number of vibrations (left panels) and mean vibration numbers on the cavity state and the local exciton state (right panels) for (a) $\omega_{\rm{c}}=1.7864$ and $2eV$, (b) $\omega_{\rm{c}}=1$ and $1.5 eV$. Other parameters: $N=100$, $\omega_0=0.17 eV$, $\varepsilon_0=2 eV$, $J/\omega_0=-1/4$, and $\lambda=1$.}
\label{FIG4}
\end{figure}

\par The left panels of Fig.~\ref{FIG4}(a) and \ref{FIG4}(b) show the total mean number of vibrations $N_{\rm{v}}$ in $|\psi_{\rm{LPP}}\rangle$ for $\omega_{\rm{c}}=1.7864$ and $2eV$, and $\omega_{\rm{c}}=1$ and $1.5 eV$, respectively. In contrast to the monotonic decreasing of $N_{\rm{v}}$ for the resonant and blue-detuned cases, a red-detuned cavity induces an increase of $N_{\rm{v}}$ from the weak- to intermediate- coupling region. As the exciton-cavity coupling is increased further up to the strong coupling regime, $N_{\rm{v}}$ will decrease from its maximal value at a crossover coupling strength. The decrease of  $N_{\rm{v}}$ with increasing $g$ indicates that the overall vibrational dressing in $|\psi_{\rm{LPP}}\rangle$ tends to fade away in the strong-coupling regime.

\par In the right panels of Fig.~\ref{FIG4}, we plot the corresponding results for the mean vibration numbers on the cavity state and on the local exciton state. For all cases considered, the mean number of vibrations $N^{(\rm{c})}_{\rm{v}}$ on the cavity state increases monotonically with increasing $g\sqrt{N}/\omega_0$, indicating an enhancement of vibrational dressing of the cavity mode. This in turn leads to a drop of $N^{(\rm{site})}_{\rm{v}}$ in the strong-coupling regime, implying an exciton-cavity coupling induced vibrational decoupling of excitons. It is interesting to note that the total mean vibration number $N_{\rm{v}}$ shows a similar trend as $N^{(\rm{site})}_{\rm{v}}$ for both the blue-detuned and red-detuned cases.
\subsection{Thermal effects}

\par As mentioned in Sec.~\ref{I}, the temperature-dependent variational Merrifield transformation method also allows us to study static properties of the system at thermal equilibrium. To calculate the thermal average of observable $\hat{O}$ at inverse temperature $\beta=1/k_{\rm{B}}T$, we turn to the Merrifield frame where the zero-order density matrix $\tilde{\rho}(\beta)$ is both separable and diagonal,
\begin{eqnarray}\label{rho_th}
&&\tilde{\rho}(\beta)\approx\tilde{\rho}_0(\beta) =\frac{1}{Z_{\rm{S}}Z_{\rm{v}}}(\sum_\eta e^{-\beta E_\eta}|\eta\rangle\langle \eta|)e^{-\beta H_{\rm{v}}},
\end{eqnarray}
where $Z_{\rm{S}}$ and $Z_{\rm{v}}=1/(1-e^{-\beta\omega_0})^N$ are the partition functions for the exciton-cavity system and the vibrational bath, respectively. The thermal average of $\hat{O}$ can thus be calculated by
\begin{eqnarray}
O(\beta)=\rm{Tr}_{\rm{S}}\rm{Tr}_{\rm{v}}[\tilde{\rho}(\beta) e^{\mathcal{S}}\hat{O}e^{-\mathcal{S}}].
\end{eqnarray}
After a straightforward calculation, we arrive at the following expressions for the finite-temperature $0-0$ and $0-1$ photoluminescence line strength (see Appendix \ref{AppB})
\begin{eqnarray}\label{I00_T}
&&I^{0-0}=\frac{\Theta^2_0}{Z_{\rm{S}}Z_{\rm{v}}} \sum_{ n}\sum_\eta e^{-\beta E_\eta} x^2_\eta e^{ik(\eta)n}  e^{-\frac{ G_n}{N}e^{-\beta\omega_0}},
\end{eqnarray}
and
\begin{eqnarray}\label{I01_T}
I^{0-1}&=&\frac{\Theta^2_0}{Z_{\rm{S}}Z_{\rm{v}}} \sum_{ n}\sum_\eta e^{-\beta E_\eta} x^2_\eta e^{ik(\eta)n} e^{- \frac{G_n }{N}e^{-\beta\omega_0}}\nonumber\\
&& e^{-\beta\omega_0 }\left[N+ \frac{2 }{N}(G_n\coth\beta\omega_0-G_0)\right].
\end{eqnarray}
The finite-temperature extension of the mean number of vibrations on the cavity state and on the local exciton state has simple forms,
\begin{eqnarray}\label{NvcT}
&&N^{(\rm{c})}_{\rm{v}}=\frac{Z_{\rm{y}}}{Z_{\rm{S}}}N(\bar{n}+h^2),
\end{eqnarray}
and
\begin{eqnarray}\label{NvnT}
&&N^{(\rm{site})}_{\rm{v}}=\frac{1}{N}\frac{Z_{\rm{x}}}{Z_{\rm{S}}} (N\bar{n}+\frac{G_0}{N}),
\end{eqnarray}
where $Z_{\rm{x}}$ and $Z_{\rm{y}}$ are defined in Eq.~(\ref{Zxy}), and $\bar{n}=1/(e^{\beta\omega_0}-1)$ is the mean occupation number of the free vibrational bath. In the zero-temperature limit, we have $Z_{\rm{x}}/Z_{\rm{S}}\to S^2$, $Z_{\rm{y}}/Z_{\rm{S}}\to C^2$, and $\bar{n}\to0$, and hence the zero-temperature results given by Eqs.~(\ref{Nv_re})--(\ref{Nvsite}) are recovered. Actually, the factor $\sum_\eta e^{-\beta E_\eta}x^2_\eta e^{ik(\eta)n}/Z_{\rm{S}}$ in $I^{0-0}$ and $I^{0-1}$, and the ratios $Z_{\rm{x/y}}/Z_{\rm{S}}$ in $N^{(\rm{c})}_{\rm{v}}$ and $N^{(\rm{site})}_{\rm{v}}$, are determined by the energy gap between the LPP state and the lowest excitonic dark state with wave vector $|k=2\pi/N\rangle$, namely, $\Delta E=E_{2\pi/N}-E_{\rm{D}}$, which is much higher than the thermal energy $k_{\rm{B}}T$ for relatively strong exciton-cavity coupling $g\sqrt{N}/\omega_0\geq 1$. We thus expect that the variational parameters $\{\tilde{f}_q\}$ and $h$ are close to those in the zero-temperature limit and almost temperature independent.
\begin{figure}
(a)\includegraphics[width=.45\textwidth]{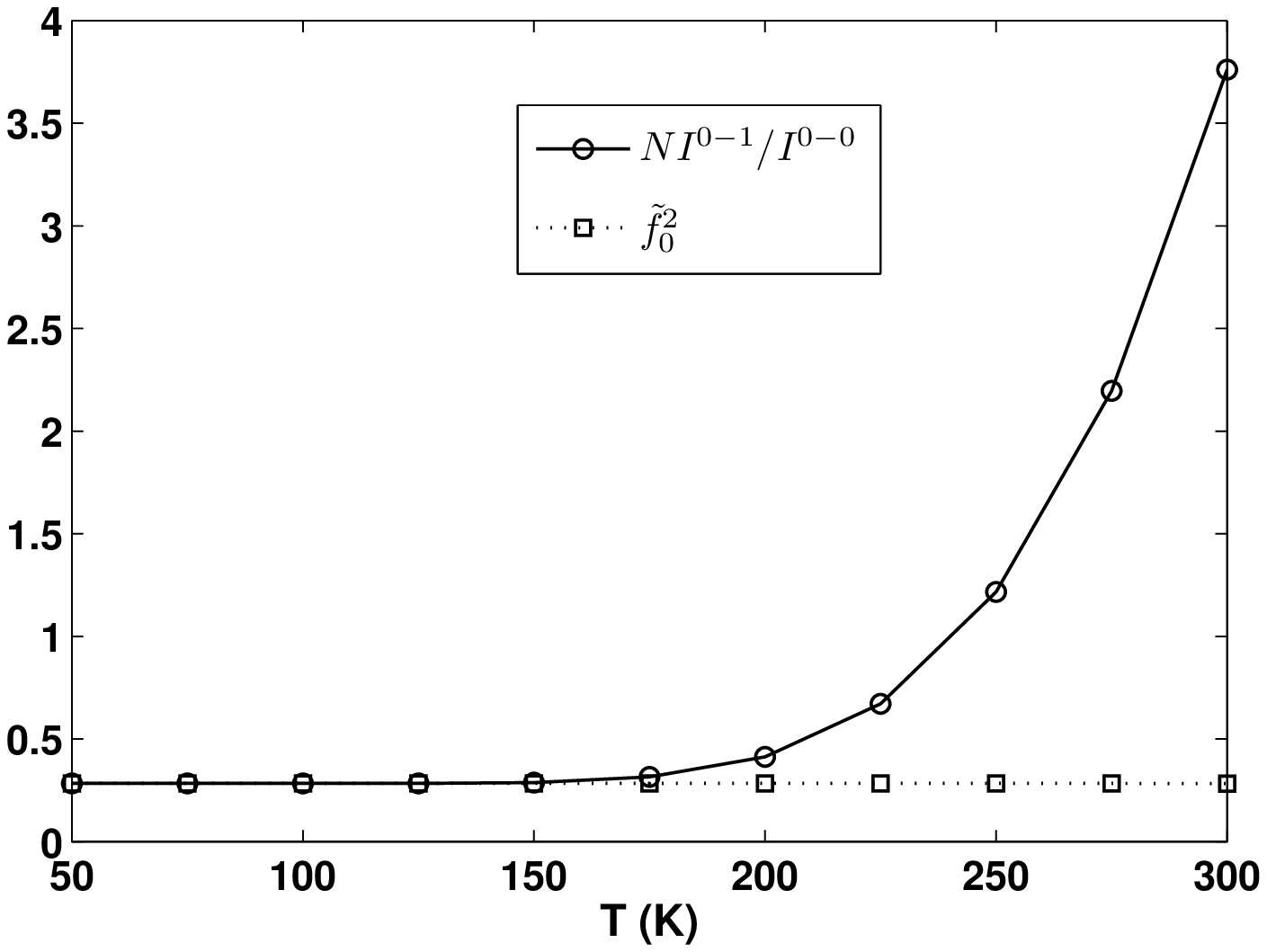}
(b)\includegraphics[width=.45\textwidth]{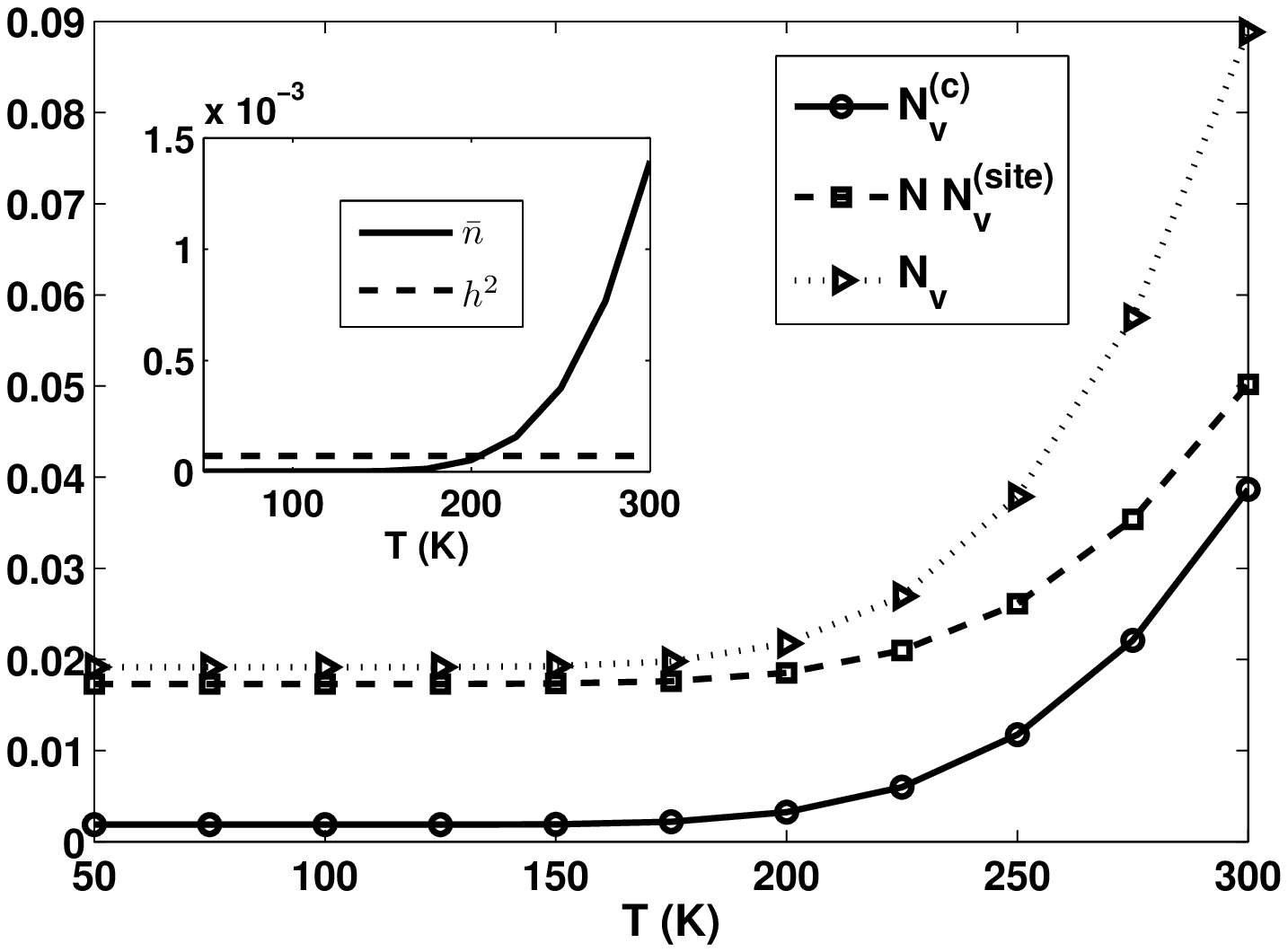}
  \caption{(a) The line-strength ratio $NI^{0-1}/I^{0-0}$. (b) The total mean number of vibrations $N_{\rm{v}}$, $N^{(\rm{c})}_{\rm{v}}$, and $NN^{(\rm{site})}_{\rm{v}}$  as functions of temperature $T$ for fixed exciton-cavity coupling $g\sqrt{N}/\omega_0=4$. The dotted curve in (a) represents the value of $NI^{0-1}/I^{0-1}$ in the zero-temperature limit, $\tilde{f}^2_0$. The inset in (b) shows the temperature dependence of the vibration occupation number $\bar{n}$ and $h^2$. Other parameters: $N=50$, $\omega_0=0.17 eV$, $\varepsilon_0=2 eV$, $J/\omega_0=-1/4$, $\lambda=1$, and $\omega_{\rm{c}}=1.7864 eV$.}
\label{FIGT}
\end{figure}

\par Fig.~\ref{FIGT}(a) shows the line-strength ratio $NI^{0-1}/I^{0-0}$ as a function of temperature $T$ for fixed exciton-cavity coupling $g\sqrt{N}/\omega_0=4$. At low temperatures, the ratio approaches the zero-temperature result, $\tilde{f}^2_0$. As the temperature increases, this ratio increases due to the decrease of $I^{0-0}$ and increase of $I^{0-1}$. This temperature dependence of $I^{0-0}$ and $I^{0-1}$ originates from the reduction of the LPP population and thermal excitation of vibrations at high temperatures. Figure~\ref{FIGT}(b) shows the temperature dependence of the mean number of vibrations. We see that $N^{(\rm{c})}_{\rm{v}}$, $N^{(\rm{site})}_{\rm{v}}$, and $N_{\rm{v}}$ all increase with increasing temperature. At low temperatures, the mean occupation number $\bar{n}$ is much smaller than $h^2$, so that $N^{(\rm{c})}_{\rm{v}}\approx NC^2h^2$. As the temperature is increased across a turning point at which $\bar{n}$ is comparable with $h^2$ [the inset of Fig.~\ref{FIGT}(b)], the thermal occupation of vibrations dominates and $N^{(\rm{c})}_{\rm{v}}$ increases rapidly to $N^{(\rm{c})}_{\rm{v}}\approx NC^2\bar{n}$ in the high-temperature limit.

\par Thus, due to the large energy gap formed in the strong exciton-cavity coupling region, the static properties of the system in thermal equilibrium are almost temperature independent below a crossover temperature which is related to the degree of vibrational dressing of the cavity and excitons. However, thermal excitation of vibrations dominates the behavior above the crossover temperature.
\section{Conclusions}\label{V}
\par In this work, we developed a microscopic theory for describing organic molecules coupled to a single cavity mode. The molecule is modeled by the Holstein Hamiltonian that explicitly includes the intramolecular vibrations. By employing a temperature-dependent variational approach combining a generalized Merrifield transformation with the Bogoliubov inequality, we could treat the exciton-vibration coupling and exciton-cavity coupling on an equal footing. The generalized canonical transformation we proposed takes the vibrational dressing of both the excitons and the cavity into account. The ground state of the system (within the single-excitation subspace), which we refer to as a lower polaron polariton, is shown to be a hybrid state of excitonic, photonic, and vibrational degrees of freedom, and contains both polaronlike and polaritonlike structures.

\par Using the above results, explicit expressions for the quasiparticle weight, the photoluminescence line strength, and the mean number of vibrations are obtained in terms of the optimal variational parameters. The dependence of these quantities upon the exciton-cavity coupling strength shows that the cavity state gains a profound vibrational dressing in the strong cavity coupling regime, while the excitons tend to decouple from the vibrations. Finally, we study the temperature dependence of the photoluminescence line strength and mean number of vibrations and show that these quantities are not affected by the temperature at relatively low temperatures, but mainly controlled by the thermal excitation of vibrations at high temperatures.

\noindent{\bf Acknowledgements:}
This work has been funded by the European Research Council (ERC-2011-AdG, Proposal No. 290981), by the European Union Seventh Framework Programme under Grant Agreement No. FP7-PEOPLE-2013-CIG-618229, by the Spanish MINECO under Contract No. MAT2014-53432-C5-5-R, and by the “María de Maeztu” programme for Units of Excellence in R\&D (Grant No. MDM-2014-0377).

\appendix
\section{The saddle point equations}\label{AppA}
We assume $N$ is even; the analysis for odd $N$ is similar. The minimized Bogoliubov free energy given by Eq.~(\ref{BB}) can be reached at the saddle point that is determined by the stationary conditions $\partial F_{\rm{B}}/\partial f_\alpha=0$, for $\alpha=0,1,...,\frac{N}{2}$, which result in
\begin{eqnarray}
\frac{A_\alpha}{Z_{\rm{S}}}=0,
\end{eqnarray}
where
\begin{eqnarray}\label{A0}
&&A_0=2\sum_{k\neq0}e^{-\beta E_k} \left[ ( f_0- \lambda)\omega_0-2( f _0- f_{ 1})\tilde{J}\cos k\coth\frac{\beta\omega_0}{2}\right]\nonumber\\
&&+e^{-\beta E_{\rm{U}}}\{ (1+\cos\theta) \left[( f_0- \lambda) \omega_0-2( f _0- f_{ 1})\tilde{J}\coth\frac{\beta\omega_0}{2}\right]\nonumber\\
&&
+\sin\theta (f_0-h)\tilde{g}\sqrt{N}\coth\frac{\beta\omega_0}{2}\}\nonumber\\
&&+e^{-\beta E_{\rm{D}}}\{ (1-\cos\theta) \left[( f_0- \lambda)\omega_0-2( f _0- f_{ 1})\tilde{J}\coth\frac{\beta\omega_0}{2}\right]\nonumber\\
&&
-\sin\theta (f_0-h)\tilde{g}\sqrt{N}\coth\frac{\beta\omega_0}{2}\},
\end{eqnarray}
\begin{eqnarray}\label{AN}
&&A_{\frac{N}{2}}=2\sum_{k\neq0}e^{-\beta E_k} \left[ f_{\frac{N}{2}}\omega_0-2( f _{\frac{N}{2}}- f_{\frac{N}{2}- 1})\tilde{J}\cos k\coth\frac{\beta\omega_0}{2}\right]\nonumber\\
&&+e^{-\beta E_{\rm{U}}}\{ (1+\cos\theta) \left[ f_{\frac{N}{2}}\omega_0-2( f _{\frac{N}{2}}- f_{\frac{N}{2}- 1})\tilde{J}\coth\frac{\beta\omega_0}{2}\right]\nonumber\\
&&
+\sin\theta (f_{\frac{N}{2}}-h)\tilde{g}\sqrt{N}\coth\frac{\beta\omega_0}{2}\}\nonumber\\
&&+e^{-\beta E_{\rm{D}}}\{ (1-\cos\theta) \left[f_{\frac{N}{2}} \omega_0-2( f _{\frac{N}{2}}- f_{\frac{N}{2}- 1})\tilde{J}\coth\frac{\beta\omega_0}{2}\right]\nonumber\\
&&
-\sin\theta (f_{\frac{N}{2}}-h)\tilde{g}\sqrt{N}\coth\frac{\beta\omega_0}{2}\},
\end{eqnarray}
and
\begin{eqnarray}\label{An}
&&A_n=2\sum_{k\neq0}e^{-\beta E_k} \nonumber\\
&&\left[f_n\omega_0-( 2f_n-f_{n-1}-f_{n+1})\tilde{J}\cos k\coth\frac{\beta\omega_0}{2}\right]\nonumber\\
&&+e^{-\beta E_{\rm{U}}}\{\sin\theta(f_n-h)\tilde{g}\sqrt{N}\coth\frac{\beta\omega_0}{2}+(1+\cos\theta)\nonumber\\
&&
\left[f_n\omega_0- (2f_n-f_{n-1}-f_{n+1})\tilde{J}\coth\frac{\beta\omega_0}{2}\right]\}\nonumber\\
&&+e^{-\beta E_{\rm{D}}}\{ -\sin\theta(f_n-h)\tilde{g}\sqrt{N}\coth\frac{\beta\omega_0}{2}+(1-\cos\theta)\nonumber\\
&& \left[ f_n\omega_0- ( 2f_n-f_{n-1}-f_{n+1})\tilde{J}\coth\frac{\beta\omega_0}{2}\right]
\},
\end{eqnarray}
for $n=1,2,...,\frac{N}{2}-1$.
\par In addition, $0=\partial F_{\rm{B}}/\partial h$ gives
\begin{eqnarray}
\frac{A_{\rm{h}}}{Z_{\rm{S}}}=0,
\end{eqnarray}
where
\begin{eqnarray}\label{Ah}
&&A_{\rm{h}}=e^{-\beta E_{\rm{U}}}[ (1-\cos\theta)Nh\omega_0
-\sin\theta (\tilde{f}_0-Nh)\nonumber\\
&&\tilde{g}\sqrt{N}\coth\frac{\beta\omega_0}{2}]+e^{-\beta E_{\rm{D}}}[ (1+\cos\theta)Nh\omega_0\nonumber\\
&&
+\sin\theta(\tilde{f}_0-Nh)\tilde{g}\sqrt{N}\coth\frac{\beta\omega_0}{2}].
\end{eqnarray}
\par An equivalent alternative form of the saddle-point equations can be obtained through linearly combining these equations, yielding
\begin{eqnarray}\label{lb_h}
\frac{\lambda}{Nh}&=&\frac{Z_{\rm{S}}}{Z_{\rm{x}}}-\frac{\omega_0}{\tilde{g}\sqrt{N}\coth\frac{\beta\omega_0}{2}}\frac{Z_{\rm{y}}}{Z_{\rm{xy}}},
\end{eqnarray}
\begin{eqnarray}\label{lb_f0}
\frac{\tilde{f}_0}{Nh}=1-\frac{\omega_0}{\tilde{g}\sqrt{N}\coth\frac{\beta\omega_0}{2}}\frac{Z_{\rm{y}}}{Z_{\rm{xy}}},
\end{eqnarray}
and
\begin{eqnarray}\label{lb_fq}
\frac{\lambda}{\tilde{f}_q}
&=&1-\frac{\tilde{g}\sqrt{N}\coth\frac{\beta\omega_0}{2}}{\omega_0}\frac{Z_{\rm{xy}}}{Z_{\rm{x}}} -2(1-\cos q)\frac{\tilde{J}\coth\frac{\beta\omega_0}{2}}{\omega_0}\nonumber\\
&&\left[1-\frac{1}{Z_{\rm{x}}}\sum_{k\neq0}e^{-\beta E_k}(1-\cos k)\right],
\end{eqnarray}
for $q\neq0$. Here,
\begin{eqnarray}\label{Zxy}
Z_{\rm{x}}&=&\sum_\eta x^2_\eta e^{-\beta E_\eta},\nonumber\\
Z_{\rm{y}}&=&\sum_\eta y^2_\eta e^{-\beta E_\eta},\nonumber\\
Z_{\rm{xy}}&=&\sum_\eta x_\eta y_\eta e^{-\beta E_\eta}.
\end{eqnarray}
\par Note that $\tilde{f}_q=\tilde{f}_{-q}$ is real, and Eqs.~(\ref{lb_h})--(\ref{lb_fq}) should also be solved self-consistently. In the zero-temperature limit, only terms related to $E_{\rm{D}}$ in Eqs.~(\ref{lb_h})--(\ref{lb_fq}) survive; we then obtain Eqs.~(\ref{lb_h_0k})--(\ref{lb_fq_0k}) in the main text.
\section{Second-order energy correction to $E_{\rm{D}}$}\label{2nd}
\par From second-order time-independent perturbation theory, the second-order corrected energy $E_{\rm{corr}}$ of the LPP state is
\begin{eqnarray}\label{ELP}
E_{\rm{corr}}&=&E_{\rm{D}}+\sum^\infty_{\nu=1}\Delta_{\nu},
\end{eqnarray}
with the $\nu$-vibration contribution
\begin{eqnarray}\label{D_nu}
&&\Delta_{\nu}= \sum_{\eta }\sum_{\sum_qn_q=\nu}\frac{|\langle \eta ;\{n_q\} |\tilde{V}  |D\rangle|^2}{E_{\rm{D}}-(E_{\eta}+\nu\omega_0)},
\end{eqnarray}
where the dressed state $|\eta;\{n_q\}\rangle=\prod_q\frac{(b^\dag_q)^{n_q}}{\sqrt{n_q!}}|\eta\rangle$ has $n_q$ vibrations in mode $q$. After a tedious but straightforward calculation, we obtain
\begin{eqnarray}\label{Delta_1}
\Delta_1&=&-4C^2\omega_0Nh^2\left(1-\frac{S^2}{1+\omega_0/(E_{\rm{U}}-E_{\rm{D}})}\right),
\end{eqnarray}
for $\nu=1$, and
\begin{eqnarray}\label{Delta_nu}
&&\Delta_\nu=\frac{1}{\nu!N^{\nu}}\sum_{\eta}\frac{1}{E_{\rm{D}}-(E_\eta+\nu\omega_0)}\nonumber\\
&&\{2(S x_{\eta}\tilde{J})^2[1+(-1)^\nu \cos(k(\eta))]\mathcal{F}_{\nu,k(\eta)}\nonumber\\
&&+2 S x_{\eta }\tilde{J}\tilde{g}\sqrt{N}[C x_{\eta } +(-1)^\nu S y_{\eta }]\mathcal{G}'_{\nu,k(\eta)}\nonumber\\
&&+(\tilde{g}\sqrt{N})^2[C x_{\eta }+(-1)^\nu S y_{\eta }]^2\mathcal{K}''_{\nu,k(\eta)}\},
\end{eqnarray}
for $\nu\geq2$, where we have defined
\begin{eqnarray}
\mathcal{F}_{\nu,p}&=&\frac{1}{N}\sum_n e^{ip n}(2G_n-G_{n-1}-G_{n+1})^\nu,\nonumber\\
\mathcal{G}'_{\nu,p}&=&\frac{1}{N}\sum_n e^{ip n}[(G'_{n}-G'_{n-1})^\nu+(G'_{n}-G'_{n+1})^\nu],\nonumber\\
\mathcal{K}''_{\nu,p}&=&\frac{1}{N}\sum_n e^{ip n}G''^\nu_n,
\end{eqnarray}
Here, $G_n$ is given by Eq.~(\ref{Gn}) and
\begin{eqnarray}\label{GGG}
G'_n=\sum_q e^{iqn}\tilde{f}_q\tilde{f}'_{-q},~G''_n=\sum_q e^{iqn}\tilde{f}'_q\tilde{f}'_{-q},
\end{eqnarray}
with
\begin{eqnarray}\label{fqprime}
\tilde{f}'_q=\tilde{f}_q-\delta_{q0}hN.
\end{eqnarray}
\section{Derivation of the finite-temperature photoluminescence line strength $I^{0-\xi}$ and the mean number of vibrations}\label{AppB}
At finite temperatures, we will adopt the zero-order thermal equilibrium density matrix for the calculation of thermal averages of observable $\hat{O}$, which is assumed to be in a separable form,
\begin{eqnarray}
\hat{O}=\hat{O}^{\rm{S}}\hat{O}^{\rm{v}},
\end{eqnarray}
where $\hat{O}^{\rm{S}}$ and $\hat{O}^{\rm{v}}$ are operators of exciton/cavity and vibrational degrees of freedom, respectively. Important examples of observables in the above form include the $0-\xi$ photoluminescence line strength $I^{0-\xi}$ (with $\hat{O}^{\rm{S}}=|0\rangle\langle0|$ and $\hat{O}^{\rm{v}}=\sum_{\sum_q n_q=\xi}|\{n_q\}\rangle\langle \{n_q\}|$) and the mean vibration number projected onto state $|\eta\rangle$ (with $\hat{O}^{\rm{S}}=|\eta\rangle\langle\eta|$ and $\hat{O}^{\rm{v}}=\sum_q b^\dag_q b_q$). It turns out to be convenient to work in the Merrifield frame where the zero-order density matrix is diagonal and separable,
\begin{eqnarray}\label{rho_th}
\tilde{\rho}(\beta)&\approx&\tilde{\rho}_0(\beta) =\tilde{\rho}_{\rm{S}}\tilde{\rho}_{\rm{v}},\nonumber\\
\tilde{\rho}_{\rm{S}}&=&\frac{1}{Z_{\rm{S}}}\sum_\eta e^{-\beta E_\eta}|\eta\rangle\langle \eta|,\nonumber\\
\tilde{\rho}_{\rm{v}}&=&\frac{1}{Z_{\rm{v}}}e^{-\beta H_{\rm{v}}}.
\end{eqnarray}
where $\beta=1/k_{\rm{B}}T$ is the inverse temperature, and $Z_{\rm{S}}$ and $Z_{\rm{v}}=1/(1-e^{-\beta\omega_0})^N$ are the partition functions for the exciton-cavity system and the vibrational bath, respectively. The representation of $\hat{O}$ in the Merrifield frame is
\begin{eqnarray}
\tilde{\hat{O}} &=&|c\rangle\langle \rm{c}|\hat{O}^{\rm{S}}_{\rm{cc}} e^{-B_{\rm{c}}}\hat{O}^{\rm{v}}e^{B_{\rm{c}}}+\sum_{mn}|m\rangle\langle n|\hat{O}^{\rm{S}}_{mn} e^{-B_m}\hat{O}^{\rm{v}}e^{B_n}\nonumber\\
&&+\sum_n(|c\rangle\langle n|\hat{O}^{\rm{S}}_{\rm{c}n} e^{-B_{\rm{c}}}\hat{O}^{\rm{v}}e^{B_n}+\rm{H.c.})
\end{eqnarray}
where $\hat{O}^{\rm{S}}_{\rm{xy}}=\langle \rm{x}|\hat{O}^{\rm{S}}|\rm{y}\rangle$ for $x,y= c$ or $\{n\}$, and H.c. stands for the Hermitian conjugate. The thermal average $O(\beta)$ at inverse temperature $\beta$ then can be calculated in the Merrifield frame as
\begin{eqnarray}\label{Obeta}
O(\beta)&=&\rm{Tr}_{\rm{S}}\rm{Tr}_{\rm{v}}[\tilde{\rho}(\beta)\tilde{\hat{O}}]\nonumber\\
&=&\hat{O}^{\rm{S}}_{\rm{cc}} \tilde{\rho}_{\rm{S,cc}} \langle e^{-B_{\rm{c}}}\hat{O}^{\rm{v}}e^{B_{\rm{c}}}\rangle_{\rm{v}}\nonumber\\
&&+ \sum_{mn}\hat{O}^{\rm{S}}_{\rm{mn}} \tilde{\rho}_{\rm{S,nm}} \langle e^{-B_{m}}\hat{O}^{\rm{v}}e^{B_{n}}\rangle_{\rm{v}}\nonumber\\
&&+  2\Re\sum_{n}\hat{O}^{\rm{S}}_{\rm{cn}} \tilde{\rho}_{\rm{S,nc}} \langle e^{-B_{\rm{c}}}\hat{O}^{\rm{v}}e^{B_{n}}\rangle_{\rm{v}}.
\end{eqnarray}
\par Let us first calculate the finite temperature $0-\xi$ photoluminescence line strength
\begin{eqnarray}
&&I^{0-\xi} = \frac{1}{Z_{\rm{S}}}\sum_{\sum_q n_q=\xi} \sum_{mn}\sum_\eta e^{-\beta E_\eta} \frac{x^2_\eta}{N}e^{ik(\eta)(n-m)}\nonumber\\
&& \langle \{n_q\}| e^{B_n}\tilde{\rho}_{\rm{v}}e^{-B_m} |\{n_q\}\rangle.
\end{eqnarray}
To calculate the matrix element in the second line of the above equation, we invoke the following two identities
 \begin{eqnarray}
&&e^{\alpha b^\dag-\alpha^*b}e^{-\sigma b^\dag b}e^{-(\gamma b^\dag-\gamma^* b)}\nonumber\\
&=&e^{-\frac{1}{2}(|\alpha|^2+|\gamma|^2)}e^{-\alpha^*\gamma e^{-\sigma}} e^{(\alpha- \gamma e^{-\sigma}) b^\dag} e^{-\beta b^\dag b} e^{( \gamma^*-\alpha^* e^{-\sigma}) b  }\nonumber\\
\end{eqnarray}
with $b$ a bosonic annihilation operator, and
\begin{eqnarray}
&&\langle \{n_q\}| e^{\sum_q\alpha_qb^\dag_q}e^{-\beta \sum_qb^\dag_q b_q}e^{\sum_q\gamma_qb_q}|\{n_q\}\rangle\nonumber\\
&=&\sum^\xi_{\chi=0} e^{-\beta(\xi-\chi)} \sum_{\sum m_q=\chi}  \prod_q\frac{(\alpha_q\gamma_q)^{m_q}}{m_q!}\frac{n_q!}{m_q!(n_q-m_q)!}.\nonumber\\
\end{eqnarray}
After some algebra, we arrive at
\begin{eqnarray}
&&I^{0-\xi} = \frac{1}{Z_{\rm{S}}Z_{\rm{v}}} \sum_{ n}\sum_\eta e^{-\beta E_\eta} x^2_\eta e^{ik(\eta)n} e^{- \sum_q\frac{|f_q|^2}{N}(1+e^{iqn-\beta\omega_0})}\nonumber\\
&&\sum^\xi_{\chi=0} e^{-\beta\omega_0(\xi-\chi)}\sum_{\sum_q n_q=\xi} \sum_{\sum m_q=\chi}\prod_q\left(\frac{|\tilde{f}_q|^2}{N}\right)^{m_q}\nonumber\\ &&\frac{(-2e^{-\beta\omega_0}+e^{iqn}e^{-2\beta\omega_0}+e^{-iqn})^{m_q}}{m_q!}\frac{n_q!}{m_q!(n_q-m_q)!}.
\end{eqnarray}
At zero temperature $T=0$, only the term with $\chi=\xi$ survives in the summation over $\chi$, and we hence recover Eq.~(\ref{Ixi}) in the main text. For large $\xi$, it is difficult to obtain $I^{0-\xi}$ in closed form. However, $I^{0-0}$ and $I^{0-1}$ can be easily calculated and we thus obtain Eqs.~(\ref{I00_T}) and (\ref{I01_T}) in the main text.
\par We next turn to the calculation of the mean number of vibrations $N^{(a)}_{\rm{v}}$ in an arbitrary exciton/cavity state $|a\rangle$. By inserting $\hat{O}^{\rm{S}}=|a\rangle\langle a|$ and $\hat{O}^{\rm{v}}=\sum_q b^\dag_q b_q$ into Eq.~(\ref{Obeta}), and using the following identity:
\begin{eqnarray}
&&\frac{1}{Z_{\rm{b}}}\rm{Tr}_{\rm{b}}[e^{-\sigma b^\dag b} e^{\alpha b^\dag-\alpha^* b} b^\dag be^{-(\gamma b^\dag-\gamma^* b)}]\nonumber\\
  &=&e^{-\frac{1}{2}(|\alpha|^2+|\gamma|^2)(1+2n_{\rm{b}}) }e^{\alpha^*\gamma(1+n_{\rm{b}})+\alpha\gamma^*n_{\rm{b}}}n_{\rm{b}}\nonumber\\
&&[1+\alpha^*\gamma( 1+n_{\rm{b}}) e^{\gamma}+\alpha\gamma^* n_{\rm{b}} - (1+n_{\rm{b}})(|\alpha|^2+|\gamma|^2)],\nonumber\\
\end{eqnarray}
with $n_{\rm{b}}=1/(e^{\sigma}-1)$, we arrive at
\begin{eqnarray}
&&N^{(a)}_{\rm{v}}= N|\langle c|a\rangle|^2  \frac{Z_{\rm{y}}}{Z_{\rm{S}}}(\bar{n}+h^2)\nonumber\\
&&+\frac{1}{N}\frac{1}{Z_{\rm{S}}}\sum_{mn}\langle m|a\rangle\langle a|n\rangle \sum_\eta e^{-\beta E_\eta} x^2_\eta e^{i k(\eta)(n-m)}\nonumber\\
&&e^{- \sum_q\frac{|\tilde{f}_q|^2}{N}[1-\cos(m-n)q](1+2\bar{n})}\sum_q e^{i\frac{|f_q|^2}{N}\sin(m-n)q}\nonumber\\
&&\left\{\bar{n}+\frac{|\tilde{f}_q|^2}{N}\left[ e^{iq(m-n)}( 1+\bar{n})^2 + e^{-iq(m-n)} \bar{n}^2 -2\bar{n}(1+\bar{n})\right]\right \}\nonumber\\
&&+\frac{1}{\sqrt{N}}\frac{1}{Z_{\rm{S}}}2\Re\sum_n \langle c|a\rangle\langle a|n\rangle\sum_\eta e^{-\beta E_\eta}x_\eta y_\eta e^{ik(\eta)n}\nonumber\\
&&e^{-\frac{1}{2}\sum_q\frac{|\tilde{f}_q|^2}{N}(1+2\bar{n})}e^{-\frac{1}{2}h(Nh-2\tilde{f}_0)(1+2\bar{n})}\nonumber\\
&&\{\bar{n}\left[N- (1+\bar{n}) \sum_{q }\frac{|f_q|^2}{N} \right]\nonumber\\
&&+ h[ f_0 ( 1+\bar{n})^2 + f_0 \bar{n}^2-\bar{n}(1+\bar{n}) Nh ]\}.
\end{eqnarray}
where $\bar{n}=1/(e^{\beta\omega_0}-1)$ is the mean occupation number of the free vibrational bath.
\par By choosing $a=c$ and $a=n$, we obtain the mean vibration numbers on the cavity state [Eq.~(\ref{NvcT})] and on the local exciton state [Eq.~(\ref{NvnT})].

\end{document}